\documentclass[superscriptaddress, groupedaddress, twocolumn, amsmath,amssymb, aps, pra]{revtex4}
\usepackage{graphicx}
\usepackage{bm}

\def\micron{{\ \mu{\rm m}}}                 

\def\Rb87{^{87}\rm{Rb}}                 

\DeclareMathAlphabet\mathbfcal{OMS}{cmsy}{b}{n}

\newcommand{\ket}[1]{\left\lvert#1\right\rangle}

\begin{document}

\title{Visible stripe phases in spin-orbital-angular-\\momentum coupled Bose-Einstein condensates}

\author{N.~-C. Chiu}
\affiliation{Institute of Atomic and Molecular Sciences, Academia Sinica, Taipei, Taiwan 10617}
\author{Y. Kawaguchi}
\email{kawaguchi@nuap.nagoya-u.ac.jp}
\affiliation{Department of Applied Physics, Nagoya University, Nagoya, 464-8603, Japan}
\author{S.~-K. Yip}
\affiliation{Institute of Physics,Academia Sinica, Taipei, Taiwan 11529}
\affiliation{Institute of Atomic and Molecular Sciences, Academia Sinica, Taipei, Taiwan 10617}
\author{Y.~-J. Lin}
\email{linyj@gate.sinica.edu.tw}
\affiliation{Institute of Atomic and Molecular Sciences, Academia Sinica, Taipei, Taiwan 10617}

\date{\today}

\begin{abstract}
Recently, stripe phases in spin-orbit coupled Bose-Einstein
condensates (BECs) have attracted much attention since they are
identified as supersolid phases. In this paper, we exploit
experimentally reachable parameters and show theoretically that annular stripe
phases with large stripe spacing and high stripe contrast can be
achieved in spin-orbital-angular-momentum coupled (SOAMC) BECs. In
addition to using Gross-Pitaevskii numerical simulations, we develop
a variational ansatz that captures the essential interaction effects
to first order, which are not present in the ansatz employed in
previous literature. Our work should open the possibility toward
directly observing stripe phases in SOAMC BECs in experiments.
\end{abstract}

\maketitle

\section{Introduction}
The realization of synthetic gauge fields and spin-orbit coupling
(SOC) for ultracold atoms has opened new opportunities for creating
and investigating topological matters in a clean and
easy-to-manipulate
environment~\cite{Dalibard11,Galitski2013,Goldman2014,Zhai2015}. In
the SOC Bose-Einstein condensates (BEC) realized in early
works~\cite{Lin11,Wu2016,Huang2016}, the internal spin states are
coupled to the center-of-mass linear momentum of the atoms via Raman
laser dressing. There, the Raman beams transfer photon momentum to
the atoms as the spin state changes. By using a similar method with
Laguerre-Gaussian (LG) Raman beams which transfer
orbital-angular-momentum (OAM) between atomic spin states,
physicists recently demonstrated a coupling between internal spin
states and the center-of-mass
OAM~\cite{Chen2018,Chen2018a,Zhang2019}. In the following, we refer
to the former as spin-linear-momentum coupling (SLMC) and the latter
as spin-orbital-angular-momentum coupling (SOAMC)~\footnote{For
simplicity, we use SLMC for both `spin-linear-momentum coupling' and
`spin-linear-momentum coupled', and use SOAMC for both
`spin-orbital-angular-momentum coupling' and
`spin-orbital-angular-momentum coupled'.}.

The interplay between interactions and SLMC leads to interesting
quantum phases~\cite{Wang2010,Ho2011,Yip2011,Li2012,Martone2014}.
For a pseudospin $1/2$ system, by tuning the Raman coupling strength
from large to small values, the energy-versus-momentum dispersion
transforms from a single minimum to double minima (see Fig.~1b). For
the latter case, whether the atoms occupy one of the minima or both
minima is determined by a competition between inter- and
intra-species interactions, where the two species refer to atoms
which occupy the two respective minima. When the atoms occupy both
minima associated with different quasimomentum, the interference
results in density modulations in the position space, which is known
as a stripe phase. When only one of the minima is occupied, it is
the separated phase, i.e., the plane-wave phase. The stripe phase in
SLMC BECs is intriguing since it spontaneously breaks the
translational symmetry (being a solid) and the $U(1)$ gauge symmetry
(being a superfluid) simultaneously, leading to a so-called
supersolid~\cite{Boninsegni2012}. Analogously, the ground state of
SOAMC BECs also has an annular stripe phase and separated phases,
which are theoretically studied in
Refs.~\cite{Qu2015,Sun2015,DeMarco2015,Chen2016}. The annular stripe
phase of SOAMC BECs corresponds to occupying both energy minima with
different quasiangular-momentum. The stripe spatial period is then
$\approx 2\pi R/\Delta \ell$, where $R$ is a typical length scale
smaller than the BEC size $R_{\rm BEC}$, and $\Delta \ell$ is the
transferred OAM between spin states in units of $\hbar$. Since $R$
is the order of micrometers, the spatial period can be made larger
than that in SLMC, which is $\lambda/2$ with $\lambda$ being the
optical wavelength of the Raman laser. The submicron stripe period
of SLMC BECs is difficult to resolve even with the state-of-the-art
quantum gas microscope~\cite{Bakr2009}.

Due to both the small spatial period and small contrast resulting
from small miscibility, direct observations of stripe phases in
position space remain elusive to date. Recently, detecting the
stripe density modulation using Bragg spectroscopy are demonstrated
in spin-linear-momentum-coupled BECs~\cite{Putra2020,Li2017}. In the
experiment with Raman-coupled internal spin states~\cite{Putra2020},
the spatial phase coherence of both the stripe and separated phase
is demonstrated interferometrically. In Ref.~\cite{Li2017}, atoms
localized within each side of a double well serve as two pseudospin
states. This circumvents the problem of detuning noises owing to the
magnetic field noises for internal spin states, and enhances the
miscibility. The observed stripe contrast is $\sim~8~\%$ limited by
heatings from the Raman driving fields which create SLMC.

\begin{figure*}
    \centering
    \includegraphics[width=7 in]{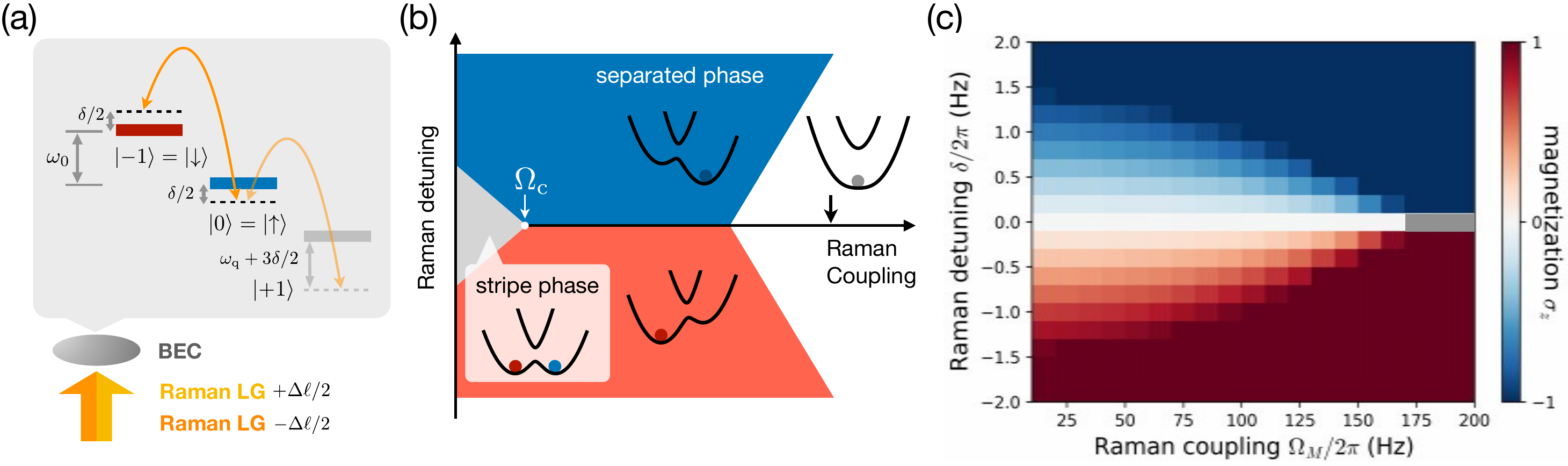}
    \caption{(a) Schematic figure of the $F=1$ SOAMC BEC with the level
    diagram. The effective two-level scheme can be realized by using a
    large quadratic Zeeman energy $\omega_q$ such that $\ket{m_F=1}$ is
    off-resonance.
    (b) Schematic SLMC phase diagram (c) SOAMC phase diagram
    of atoms in a harmoinc trap. $\Delta \ell=20, r_M=5\micron,
    R_{\rm TF}=12.5\micron$, and $\mu/h=926$~Hz.
    For $\delta=0$, the ground state is the stripe (separated) phase with
    $\Omega_M<(>)\Omega_c$. For small nonzero detuning, the critical
    coupling is smaller than the $\Omega_c(\delta=0)$ and decreases with
    increasing $|\delta|$. Near $\Omega_M=0$, the stripe phase exists
    for $|\delta|/2\pi \lesssim 1$~Hz. At $\delta=0$, the magnetization
    $\sigma_z=N_{\uparrow}-N_{\downarrow}$ is zero; the stripe contrast
    at $\delta=0$ is larger than that at any other nonzero $\delta$. The grey bar at
    $\delta=0$ and $\Omega_M>\Omega_c$ indicates the separated phase can have either $\sigma_z=1$
    or $-1$.}
\end{figure*}

In this paper, we exploit the advantages of SOAMC systems and
demonstrate the feasibility to directly observe annular stripe
phases in situ with practical experimental parameters. We observe
that interactions reduce the stripe density contrast. Here the
interaction strength is $\varepsilon_{\rm int}/E_L <1$, where
$\varepsilon_{\rm int}$ is the mean field interaction energy and
$E_L=\hbar^2\Delta \ell^2/2mR^2$ is the characteristic energy scale
of SOAMC systems. The effects of interactions discussed in previous
papers~\cite{Li2012,Sun2015,Chen2016,Chen2019} are based on the wave
function ansatz that is not fully self-consistent in the presence of
interaction.  Even within first order in interaction strength, we
find that the results of
Refs.~\cite{Li2012,Sun2015,Chen2016,Chen2019} are subject to
significant corrections. We use an improved ansatz and obtain
results that are correct to first order in interaction. While in
SLMC systems, the analogous interaction strength is
$\varepsilon_{\rm int}/4E_r$ and is typically small, where the
photon recoil energy $E_r$ is larger than $E_L$. We investigate how
the stripe density contrast depends on experimentally accessible
parameters: the transferred angular momentum $\Delta \ell$, the size
of the OAM-carrying LG Raman beam, the BEC cloud size, and the mean
field energy. By optimizing these parameters, we achieve a stripe
period of $\sim 2\micron$ at most and a $\lesssim 30~\%$ contrast of
density modulations. This is detectable using high-resolution
imaging with about $1\micron$ resolutions~\cite{Bakr2009} . Further,
the contrast can be made larger than $30~\%$ by increasing the BEC
cloud size. Finally, we point out that by using synthetic clock
states~\cite{Trypogeorgos2018}, the stripe phase of the
thermodynamic ground state can be stable against external magnetic
field noises despite the narrow detuning window within which the
stripe phase exits.

\section{Formalism}
We consider pseudospin $1/2$ atoms tightly confined along $z$ in a
quasi-2D geometry, where $\hbar \omega_z > \mu$ with $\omega_z$
being the trap frequency along $z$ and $\mu$ the chemical potential.
Two Raman beams couple the two spin states with a transfer of
orbital-angular-momentum (OAM) $\Delta \ell$ in unit of $\hbar$, and
the frequency difference between the two beams is $\Delta \omega_L$.
In the rotating frame at frequency $\Delta \omega_L$ with rotating
wave approximation, the single-particle Hamiltonian is
\begin{align}\label{eq:H0}
\hat{H}_0=& \left[\frac{-\hbar^2}{2m}\frac{\partial}{r\partial r}(r
\frac{\partial}{\partial r})+\frac{L_z^2}{2m r^2}+V(r) \right]
\otimes {\hat 1} \\ 
&+\frac{\hbar \delta}{2} \Hat{\sigma}_z+\frac{\hbar
\Omega(r)}{2}\left[ \cos (\Delta \ell \phi) \Hat{\sigma}_x- \sin
(\Delta \ell \phi) \Hat{\sigma}_y\right], \nonumber
\end{align}
where $L_z=-i\hbar \partial_\phi$ is the angular momentum operator,
$V(r)$ is the spin-independent trapping potential, $\delta=\Delta
\omega_L-\omega_0$ is the Raman detuning, and
$\hbar\omega_0=E_{\downarrow}-E_{\uparrow}$ is the energy splitting
between $\ket{\downarrow}$ and $\ket{\uparrow}$. The Raman beams are
two Laguerre-Gaussian beams of order $\Delta \ell/2$ and $-\Delta
\ell/2$, and the coupling strength is
\begin{align}
\Omega(r)=e^{\Delta \ell/2}\Omega_M
\left(\frac{r}{r_M}\right)^{\Delta \ell}\exp\left[-\frac{\Delta
\ell}{2} \frac{r^2}{r_M^2}\right],
\end{align}
where the peak coupling $\Omega_M$ is at $r=r_M$, and the waist of
each beam is $w=2 r_M/\sqrt{\Delta \ell}$.

\begin{figure*}
    \centering
    \includegraphics[width=6 in]{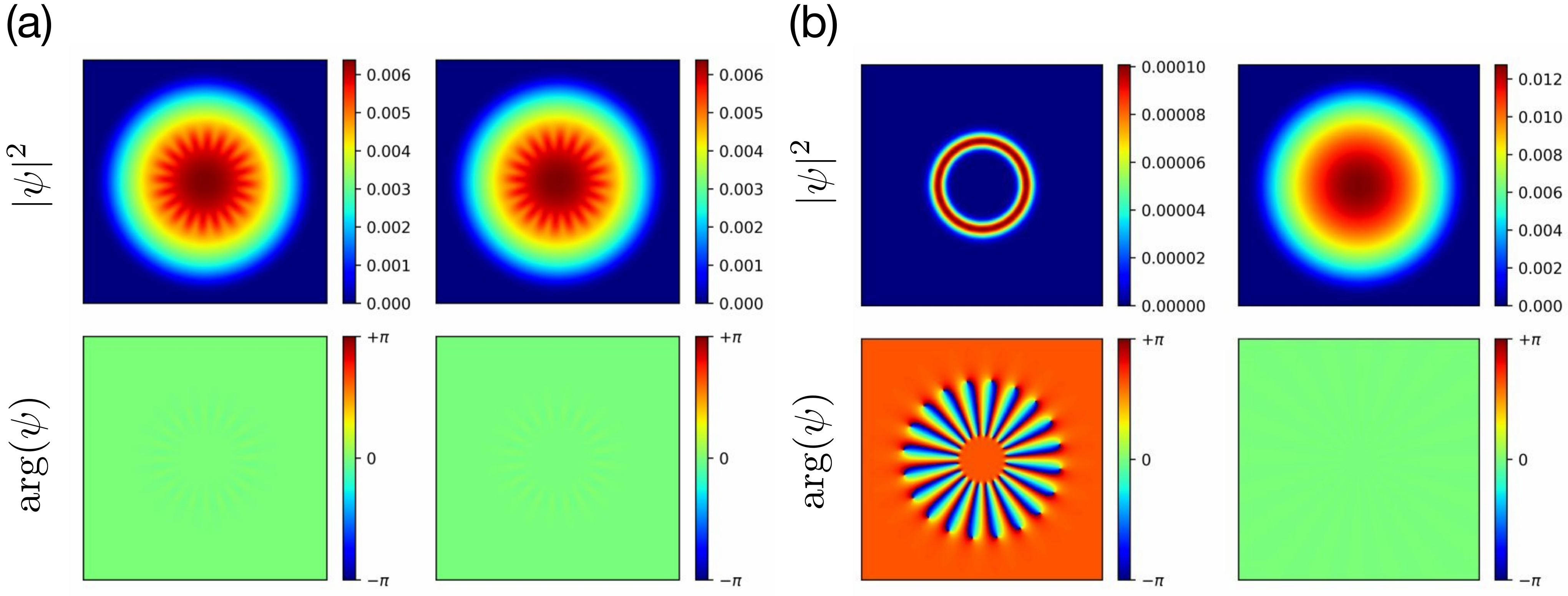}
    \caption{Images of the stripe phase at $\Omega_M/2\pi=150$~Hz
    $<\Omega_c$ in (a), and the separated phase at $\Omega_M/2\pi=200$~Hz $>\Omega_c$ in (b); left (right)
    panels indicate $\ket{\uparrow}(\ket{\downarrow})$, and top (bottom)
    panels indicate the density (phase of the wave function). $\delta=0$ and other parameters
    are given in Fig.~1c. The image scale is $28\micron \times 28\micron$.}
\end{figure*}

In addition to $\hat{H}_0$, we have the mean field energy
\begin{align}\label{eq:Eint}
E_{\rm int}=\int d^3 r \left(
\frac{g_{\uparrow\uparrow}}{2}|\psi_\uparrow|^4+
\frac{g_{\downarrow\downarrow}}{2}|\psi_\downarrow|^4
+g_{\uparrow\downarrow}|\psi_\uparrow|^2|\psi_\downarrow|^2\right),
\end{align}
where $|\psi_{\uparrow}|^2,|\psi_{\downarrow}|^2$ are the 2D density
of $\ket{\uparrow},\ket{\downarrow}$. The wave functions are
normalized as $\int dr r \int d\phi n(r,\phi)=N$ where
$n=|\psi_{\uparrow}|^2+|\psi_{\downarrow}|^2$ and $N$ is the number
of atoms. The 2D interaction strengths are
$g=g_{\uparrow\uparrow}=g_{\downarrow\downarrow}$ and
$g_{\uparrow\downarrow}$. We define $g_1=(g+g_{\uparrow
\downarrow})/2,g_2=(g-g_{\uparrow \downarrow})/2$, $g_2$ being the
spin-dependent interaction strength. We use real experimental
parameters by taking the pseudospin states as
$\ket{\uparrow}=|F=1,m_F=0\rangle$ and
$\ket{\downarrow}=|1,-1\rangle$ of $^{87}$Rb atoms, for which
$g=(g_{00}+g_{-1,-1})/2,g_{\uparrow \downarrow}=g_{0,-1}$ with
$g_{00}=4 \pi \hbar^2 a_{00}/(m\sqrt{2\pi}R_z),g_{-1,-1}=g_{0,-1}=4
\pi \hbar^2 a_{-1,-1}/(m\sqrt{2\pi}R_z)$, and
$R_z=\sqrt{\hbar/m\omega_z}$. The scattering lengths are
$a_{00}=100.86 a_B$ and $a_{-1,-1}=100.40 a_B$, where $a_B$ is the
Bohr radius~\cite{VanKempen2002}. This gives $g>g_{\uparrow
\downarrow}$ and positive $g_2/g_1=0.00114$. As compared to the
realistic case with $g_{\uparrow\uparrow} \neq
g_{\downarrow\downarrow}$, here our simplification of using
$g=g_{\uparrow\uparrow}=g_{\downarrow\downarrow}$ is based on the
results of uniform SLMC systems in the absence of trapping
potentials in Ref.~\cite{Li2012}, which is just a shift in detuning
for the ground state. We show that this is a good approximation for
the trapped atoms with inhomogeneous $n(r)$ under SOAMC.

For $\delta=0$ in the non-interacting limit, the ground state may be
expressed as

\begin{align}\label{eq:spinor_Psi}
\left(\begin{array}{c} \psi_{\uparrow}\\
\psi_{\downarrow}
\end{array}\right)= \sqrt{\bar{n}(r)}[&
C_+ e^{i \frac{\Delta \ell}{2}\phi}\left(
\begin{array}{c}
\sin \theta(r)e^{i\frac{\Delta \ell}{2}\phi} \\
-\cos \theta(r)e^{-i\frac{\Delta \ell}{2}\phi}
\end{array}\right)\nonumber \\+&
C_- e^{-i\frac{\Delta \ell}{2}\phi}\left(
\begin{array}{c}
\cos \theta(r)e^{i\frac{\Delta \ell}{2}\phi}\\
-\sin \theta(r)e^{-i\frac{\Delta \ell}{2}\phi}
\end{array}\right)].
\end{align}

where $\ell>0$, $|C_+|^2+|C_-|^2=1$, and $\bar{n}(r)$ is the density
after azimuthal average with $\int dr 2\pi r\bar{n}=N$. Since the
Raman coupling has a phase winding number $\Delta \ell$, i.e., an
OAM of light, the Raman beams couple
$\ket{\uparrow,\ell_{\uparrow}}$ to
$\ket{\downarrow,\ell_{\downarrow}}$ where the OAM difference
between the spin states is $\ell_{\uparrow}-\ell_{\downarrow}=\Delta
\ell$. By introducing the quasiangular momentum $\ell$,
$\ell_\uparrow$ and $\ell_\downarrow$ are rewritten as
$\ell_{\uparrow}=\ell+\Delta \ell/2$ and
$\ell_{\downarrow}=\ell-\Delta \ell/2$.
Then, Eq.~\eqref{eq:spinor_Psi} is referred as
`two-quasiangular-momentum ansatz', which have two running wave
components along $\phi$ with quasiangular momentum $\pm\ell$. For
sufficiently small Raman coupling, the ground state has $\ell=\Delta
\ell/2$~\cite{Qu2015,Sun2015}, and we focus on this regime
throughout this paper. With $\ell=\Delta \ell/2$, there is a
critical coupling $\Omega_c$ below which the ground state has
$|C_+||C_-|>0$. For $\Omega_M<\Omega_c$, Eq.~\eqref{eq:spinor_Psi}
shows that the spin component $\ket{\uparrow}$ ($\ket{\downarrow}$)
has an OAM superposition of $\ell_{\uparrow}=\Delta \ell$ and 0
($\ell_{\downarrow}=0$ and $-\Delta \ell$), leading to a density
modulation along $\phi$, which is then called stripe phase (see
Fig.~2a). Here,

\begin{align}\label{eq:density_stripe}
|\psi_{\uparrow}|^2/\bar{n} =& |C_+|^2 \sin^2
\theta+|C_-|^2\cos^2 \theta \nonumber \\ 
& + |C_+||C_-|\sin 2\theta \cos (\Delta \ell \phi +\varphi), \nonumber \\
|\psi_{\downarrow}|^2/\bar{n} =& |C_+|^2 \cos^2 \theta+|C_-|^2\sin^2 \theta \nonumber \\ 
& +|C_+||C_-|\sin 2\theta \cos (\Delta \ell \phi +\varphi), \nonumber \\
n =& \bar{n}\left[1+2|C_+||C_-|\sin 2\theta \cos (\Delta \ell \phi
+\varphi) \right],
\end{align}

where $\varphi$ is the relative phase between $C_+$ and $C_-$. With
$\Omega_M>\Omega_c$, the ground state is the separated phase with
$|C_+||C_-|=0$, i.e., $|C_+|=1,|C_-|=0$ or $|C_-|=1,|C_+|=0$  (see
Fig.~2b), which are equivalent for $\delta=0$. For the stripe phase
with $|C_+||C_-|>0$, the ground state has $|C_+|^2=|C_-|^2=1/2$ and
$|C_+|=|C_-|$ where $|C_+|^2|C_-|^2$ is maximized. Note that at
$|C_+|=|C_-|$ the wave function Eq.~\eqref{eq:spinor_Psi} is an
eigenstate of the time-reversal operator $T=\hat{\sigma}_x K$ with
$K$ being the complex-conjugate operator, which is possible because
the Hamiltonian commutes with $T$. At radial position $r$, the
contrast of the azimuthal density modulation is
\begin{align}\label{eq:nc_max_min}
\eta(r)=\frac{n_{\rm max}(r)-n_{\rm min}(r)}{2n_{\rm avg}(r)},
\end{align}
where $n_{\rm max},n_{\rm min}$ and $n_{\rm avg}$ are the maximum,
minimum and average of the density along $\phi$, respectively. Then,
the contrast of both $\ket{\uparrow}$ and $\ket{\downarrow}$ from
Eq.~\eqref{eq:density_stripe} is
\begin{align}\label{eq:nc_sin}
\eta(r)= \sin 2\theta(r),
\end{align}
and the spatial period of the density stripe is $2\pi r/\Delta
\ell$.

Now we consider the general form of the spinor wave function for all
interaction strength, which is
\begin{align}\label{eq:spinor_Psi_Apm_all}
\left(\begin{array}{c} \psi_{\uparrow}\\
\psi_{\downarrow}
\end{array}\right)= \sum_{p}\left(
\begin{array}{c}
a_{-\frac{\Delta \ell}{2}+p \Delta \ell}\\
b_{-\frac{\Delta \ell}{2}+p \Delta \ell}
\end{array}\right)e^{i (-\frac{\Delta \ell}{2}+p \Delta \ell)\phi},
\end{align}
, where $p=0,\pm1,...$ are integers. The multiple OAM components
differing by $\Delta \ell$ are due to the nonlinear interaction
term, and Eq.~\eqref{eq:spinor_Psi_Apm_all} is similar to that in
Ref.~\cite{Li2013} for spin-linear-momentum coupling. In the
two-quasiangular-momentum ansatz Eq.~\eqref{eq:spinor_Psi} with
$\ell=\Delta \ell/2$, it has only $a_0,a_{\Delta \ell}$ and
$b_0,b_{-\Delta \ell}$. For small $\theta$, and small interactions
in the first order perturbation regime,
Eq.~\eqref{eq:spinor_Psi_Apm_all} can be simplified as (see
appendix) having only $a_0,a_{\pm\Delta \ell}$ and $b_0,b_{\pm\Delta
\ell}$,
\begin{align}\label{eq:spinor_Psi_Apm}
\left(\begin{array}{c} \psi_{\uparrow}\\
\psi_{\downarrow}
\end{array}\right)= \sqrt{\bar{n}(r)}[&A_+ e^{i \frac{3\Delta \ell}{2}\phi}\left(
\begin{array}{c}
0\\
e^{-i\frac{\Delta \ell}{2}\phi}
\end{array}\right) \nonumber \\
+&C_+ e^{i \frac{\Delta \ell}{2}\phi}\left(
\begin{array}{c}
\sin \theta(r)e^{i\frac{\Delta \ell}{2}\phi} \\
-\cos \theta(r)e^{-i\frac{\Delta \ell}{2}\phi}
\end{array}\right)\nonumber \\+&
C_- e^{-i\frac{\Delta \ell}{2}\phi}\left(
\begin{array}{c}
\cos \theta(r)e^{i\frac{\Delta \ell}{2}\phi}\\
-\sin \theta(r)e^{-i\frac{\Delta \ell}{2}\phi}
\end{array}\right) \nonumber \\
+& A_- e^{i -\frac{3 \Delta \ell\phi}{2}}\left(
\begin{array}{c}
e^{i\frac{\Delta \ell}{2}\phi}\\
0
\end{array}\right)].
\end{align}
This is referred as `four-quasiangular-momentum ansatz', which has
$\ell=\pm \Delta \ell/2, \pm 3\Delta \ell/2$. Since the ground state
energy is independent of the relative phase between $C_+$ and $C_-$,
we can take $C_+<0,C_->0$ as real values without loss of generality.
Similar to the case with two-quasiangular-momentum ansatz, we
consider states that are eigenstates of the time-reversal operator
$T$, which will be confirmed by numerical simulations. The
variational parameters then satisfy
\begin{align}
|C_+|=|C_-|,|A_+|=|A_-|, \alpha_+ + \alpha_-= 0~\rm{mod}~2\pi,
\end{align}
where $A_+=|A_+|e^{i \alpha_+},A_-=|A_-|e^{i \alpha_-}$. The
$\cos(\Delta \ell \phi)$ term in the density modulation is $C_- \cos
\theta (C_+\sin \theta+A_-^\ast)e^{i\Delta \ell \phi}+$ c.c. for
$\ket{\uparrow}$, and $-C_+ \cos \theta (-C_-\sin
\theta+A_+)e^{i\Delta \ell \phi}+$ c.c. for $\ket{\downarrow}$.
Thus, for the ground state with minimized density modulation, the
relative phase between $C_+\sin \theta$ and $A_-^\ast$ and between
$-C_-\sin \theta$ and $A_+$ is $\pi$. This gives real and positive
$A_+,A_-$. We then have
\begin{align}\label{eq:Cpm_Apm}
-C_+=C_- >0,A_+=A_-=A_{\pm}>0.
\end{align}
Similarly, if we choose $C_+,C_->0$, the condition becomes $C_+=C_-
>0,A_+= -A_->0$. With $C_+,C_-,A_+,A_-$ as real numbers, the
densities of $\ket{\uparrow}$ and $\ket{\downarrow}$ are
\begin{align}\label{eq:density_stripe_Apm}
|\psi_{\uparrow}|^2/\bar{n}&=C_+^2 \sin^2 \theta + C_-^2 \cos^2
\theta +A_-^2 \nonumber\\
&+2\cos \theta(C_+C_-\sin \theta+A_-C_-)\cos(\Delta \ell \phi) \nonumber\\
&+2A_-C_+ \sin \theta \cos (2\Delta \ell \phi),\nonumber\\
|\psi_{\downarrow}|^2/\bar{n}&=C_-^2 \sin^2 \theta + C_+^2 \cos^2
\theta+A_+^2 \nonumber\\ &+2\cos
\theta(C_+C_-\sin \theta-A_+C_+)\cos(\Delta \ell \phi)\nonumber\\
&-2A_+C_- \sin \theta \cos (2\Delta \ell \phi).
\end{align}
The normalization is $C_+^2+C_-^2+A_+^2+A_-^2=1$, leading to
\begin{align}\label{eq:Cpmsq}
C_+^2=C_-^2=(1-A_+^2-A_-^2)/2,
\end{align}
and for small $A_+,A_-$, $C_+^2=C_-^2 \lesssim 1/2$ and $C_{\pm}^2
\sin^2 \theta + C_{\mp}^2 \cos^2 \theta \approx 1/2$. Thus, the
density contrast of the $\cos(\Delta \ell \phi)$ term in
Eq.~\eqref{eq:density_stripe_Apm} is
\begin{align}
\eta_{\uparrow} &\approx \left|2 \sin 2\theta C_+C_- +4\cos \theta A_-C_-\right|,\nonumber\\
\eta_{\downarrow} &\approx \left|2 \sin 2\theta C_+C_- -4\cos \theta
A_+C_+\right|,
\end{align}
by using Eq.~\eqref{eq:nc_max_min} with $(n_{\rm max}-n_{\rm
min})/2$ equal to the amplitude of the $\cos(\Delta \ell \phi)$ term
and $n_{\rm avg}\approx 1/2$. With Eq.~\eqref{eq:Cpm_Apm}, the
contrast is
\begin{align}\label{eq:contrast_Apm}
\eta_{\uparrow}=\eta_{\downarrow}\approx  \left|\sin 2\theta -
2\sqrt{2}\cos \theta A_{\pm}\right|.
\end{align}
Since the density modulation of the $\cos(2\Delta \ell \phi)$ term
is much smaller than $\eta_{\uparrow,\downarrow}$, we use
Eq.~\eqref{eq:contrast_Apm} as the contrast in our simulations.

\section{Simulations methods}
We perform both the Gross-Pitaevskii (GP) simulations and the
variational calculations to find the ground state in the SOAMC
system. The GP simulation gives the ground state with the full
Hamiltonian including both $\hat{H_0}$ and the interaction energy.
Additionally, we perform the variational calculations with a
simplified picture: we neglect the radial kinetic energy associated
with $\partial_r$ in $\hat{H_0}$, thus the rest of all the energy
terms are functions of radial position $r$. We find the results of
GP and variational methods have good agreements.

\subsection{Gross-Pitaevskii ground state}
We use the GP simulations to find the ground state by numerically
solving the Gross-Pitaevskii equation (GPE). We perform imaginary
time propagations, where the initial state of the stripe phase for
the imaginary time propagation is
\begin{align}
\left(\begin{array}{c} \psi_{\uparrow}\\
\psi_{\downarrow}
\end{array}\right)= \sqrt{\frac{n_{\rm TF}(r)}{2}} \frac{1}{\sqrt{2}}\left(
\begin{array}{c}
1+e^{i \Delta \ell \phi}\\
1+e^{-i \Delta \ell \phi}
\end{array}\right),
\end{align}
and for the separated phase it is
\begin{align}
\left(\begin{array}{c} \psi_{\uparrow}\\
\psi_{\downarrow}
\end{array}\right)= \sqrt{\frac{n_{\rm TF}(r)}{2}} \left(
\begin{array}{c}
e^{i \Delta \ell \phi}\\
1
\end{array}\right).
\end{align}
The initial state of the stripe phase has a superposition of OAM
differing by $\Delta \ell$ in either spin up and down, such that all
the OAM components differing by $\Delta \ell$ [see
Eq.~\eqref{eq:spinor_Psi_Apm_all}] can be reached in the final
ground state. The initial state of the separated phase corresponds
to $C_-=1, C_+=0$. After the numerical computation, we compare the
energy differences between the two phases and determine the ground
state from the lower energy state.

\subsection{Variational Method}
We adopt a variational method to minimize the energy $E_{\rm var}$
for $\delta=0$ and obtain the variational ground state. $E_{\rm
var}$ includes the single particle Hamiltonian in Eq.~\eqref{eq:H0},
but excluding the radial kinetic energy from $\partial_r$, and the
mean field interaction Eq.~\eqref{eq:Eint}. This gives $E_{\rm
var}=\int dr r 2\pi \bar{n}(r) \varepsilon(r)$ where
$\varepsilon(r)$ is the energy per atom after the azimuthal average.

We discuss calculations based on the two-quasiangular-momentum
ansatz, Eq.~\eqref{eq:spinor_Psi}, and four-quasiangular-momentum
ansatz, Eq.~\eqref{eq:spinor_Psi_Apm}, respectively. The variational
ground state from the simple Eq.~\eqref{eq:spinor_Psi} agrees with
our GP simulation in the non-interacting limit. This variational
form, Eq.~\eqref{eq:spinor_Psi}, is used in earlier
papers~\cite{Li2012,Sun2015,Chen2016,Chen2019}; ~\cite{Li2012} for
SLMC and~\cite{Sun2015,Chen2016,Chen2019} for SOAMC BECs. In our
simulations, we find the variational ground state from
Eq.~\eqref{eq:spinor_Psi} is inconsistent with the GP result in the
non-negligible interaction regime, where the additional OAM
components must be taken into account as the ansatz
Eq.~\eqref{eq:spinor_Psi_Apm}.

\subsubsection{Two-quasiangular-momentum ansatz}
With the ansatz of Eq.~\eqref{eq:spinor_Psi}, the variational energy
per atom $\varepsilon^{\rm var0}$ is given by
\begin{subequations}\label{eq:E_variational}
\begin{align}
\varepsilon^{\rm var0}=&\varepsilon_0^{\rm var0}+\varepsilon_{\rm int}^{\rm var0},\\
\varepsilon_0^{\rm var0}=&-\frac{\hbar\Omega(r)}{2}\sin
2\theta+E_L\frac{1-\cos 2\theta}{2}\\
\varepsilon_{\rm int}^{\rm var0}=&\frac{\bar{n}(r) g_1}{2}+
\frac{\bar{n}(r) g_2}{2}\cos^2 2\theta \nonumber \\
&+\beta\left[\bar{n}(r)
g_1\sin^2 2\theta-2\bar{n}(r)g_2\cos^2 2\theta \right].
\end{align}
\end{subequations}
$\varepsilon_0^{\rm var0}$ is the single-particle energy arising
from the Raman coupling and the centrifugal potential $L_z^2/2mr^2$,
where the latter is characterized by $E_L=\hbar^2(\Delta
\ell)^2/2mr^2$ at position $r$. Here we exclude the trap energy
$V(r)$ in $\hat{H}_0$ since $V(r)$ doesn't depend on any variational
parameters and is simply an offset. $\varepsilon_{\rm int}^{\rm
var0}$ is the mean field interaction energy with $\beta=|C_+|^2
|C_-|^2$ satisfying $0\leq \beta\leq 1/4$. Given the local energy
$\varepsilon^{\rm var0}(r)$ at a radial position $r$ with the
averaged density $\bar{n}(r)$, we take $\theta(r)$ and $\beta$ as
variational parameters.
$\beta=0$ for the separated phase with $|C_+|=1,|C_-|=0$ or
$|C_+|=0,|C_-|=1$, and $\beta=1/4$ for the stripe phase with
$|C_+|=|C_-|=1/\sqrt{2}$. Within $0\leq \beta\leq 1/4$, the energy
difference between the stripe and the separated phase is lowest at
either $\beta=0$ or $\beta=1/4$ (see appendix). At a given $\beta$,
minimizing $\varepsilon^{\rm var0}(r)$ with respect to $\theta$
determines $\theta$ as a solution of the following equation:

\begin{align}\label{eq:theta_sol}
\frac{\sin 2\theta}{\cos 2\theta }=\frac{\hbar
\Omega(r)}{E_L(r)}-\frac{\bar{n}(r)g_1}{E_L(r)}\left(4\beta +8\beta
\frac{g_2}{g_1}- 2\frac{g_2}{g_1}\right)\sin 2\theta
\end{align}

In the non-interacting case, the solution of
Eq.~\eqref{eq:theta_sol} is $\tan 2\theta_{0}=\Omega/E_L$, or
equivalently
\begin{align*}
\sin 2\theta_0=\frac{\hbar\Omega/E_L}{\sqrt{1+(\hbar\Omega/E_L)^2}},
\end{align*}
which can be approximated for small $\hbar\Omega/E_L$ as
\begin{align*}
\sin 2\theta_0\approx \hbar\Omega/E_L.
\end{align*}
With interactions, the solution $\theta^{\rm var0}$ of the stripe
phase with $\beta=1/4$ is smaller than $\theta_0$, given by
\begin{align}\label{eq:theta_var0}
\sin 2\theta^{\rm var0}- \cos 2\theta^{\rm var0} \left(\frac{\hbar
\Omega(r)}{E_L(r)}-\frac{\bar{n}(r)g_1}{E_L(r)}\sin 2\theta^{\rm
var0} \right)=0,
\end{align}
from which the contrast is given by
\begin{align}\label{eq:nc_var0}
\eta^{\rm var0}(r)=\sin 2\theta^{\rm var0}(r)
\end{align}
as derived in Eq.~\eqref{eq:nc_sin}. We use $\bar{n}(r)$ obtained
from the GP simulation, which is the same for the stripe phase and
separated phase and is well approximated by the Thomas-Fermi (TF)
profile except for small $r_M$. By expanding to first order in
$\bar{n}g_1/E_L$ and $\hbar\Omega/E_L$,
\begin{align}\label{eq:theta_var0_expand}
\theta^{\rm var0} \approx \frac{\hbar\Omega}{2
E_L}(1-\frac{\bar{n}g_1}{E_L}),~~
\eta^{\rm var0} \approx
\frac{\hbar\Omega}{E_L}(1-\frac{\bar{n}g_1}{E_L}).
\end{align}
For the separated phase with $\beta=0$, $\theta_{\rm sep}$ is well
approximated with $\theta_0$ owing to $\bar{n}g_2 \ll E_L$.

\subsubsection{Four-quasiangular-momentum ansatz}
With the ansatz of Eq.~\eqref{eq:spinor_Psi_Apm}, the
single-particle part of the variational energy is given by
\begin{align}
\varepsilon_0^{\rm var} =&\left[-\frac{\hbar\Omega(r)}{2}\sin
2\theta+E_L\frac{1-\cos 2\theta}{2}\right](C_+^2+C_-^2)\nonumber \\
&+ E_L (A_+^2+A_-^2),
\end{align}
where $C_+,C_-,A_+,A_-$ are real. The interaction energy
$\varepsilon_{\rm int}^{\rm var}$ is also a function of
$\theta,C_+,C_-,A_+,A_-$; by using Eq.~\eqref{eq:Cpm_Apm} for the
stripe phase, we plug in
$C_{\pm}={\mp}\sqrt{(1-2A_{\pm}^2)/2},A_+=A_-=A_{\pm}$ and obtain
\begin{align}
\varepsilon_{\rm int}^{\rm var}=&-\frac{1}{8}\bar{n} g_1 [-5+ 8 A_{\pm} (2 - 4 A_{\pm}^2)^{3/2}
     \cos^2 \theta \sin\theta      \nonumber\\
     &+ (1 - 4 A_{\pm}^2 + 4 A_{\pm}^4) (1 -
      2 \sin^2 2\theta)  - 12 A_{\pm}^2 + 28 A_{\pm}^4 ].
\end{align}
Then we minimize $\varepsilon^{\rm var}=\varepsilon_0^{\rm
var}+\varepsilon_{\rm int}^{\rm var}$ with respect to $(\theta,
A_{\pm})$, respectively, giving the numerical solutions for the
stripe phase, $\theta^{\rm var}$ and $A_{\pm}^{\rm var}>0$, where
the sign of $A_{\pm}^{\rm var}$ agrees with Eq.~\eqref{eq:Cpm_Apm}.
The contrast of the $\cos(\Delta \ell \phi)$ term is
\begin{align}\label{eq:nc_var_Apm}
\eta^{\rm var}(r)=\left|\sin 2\theta^{\rm var}(r) -2\sqrt{2}\cos
\theta^{\rm var}(r) A_{\pm}^{\rm var}(r)\right|
\end{align}
for both $\ket{\uparrow}, \ket{\downarrow}$ following
Eq.~\eqref{eq:contrast_Apm}.
\begin{figure*}
    \centering
    \includegraphics[width=7 in]{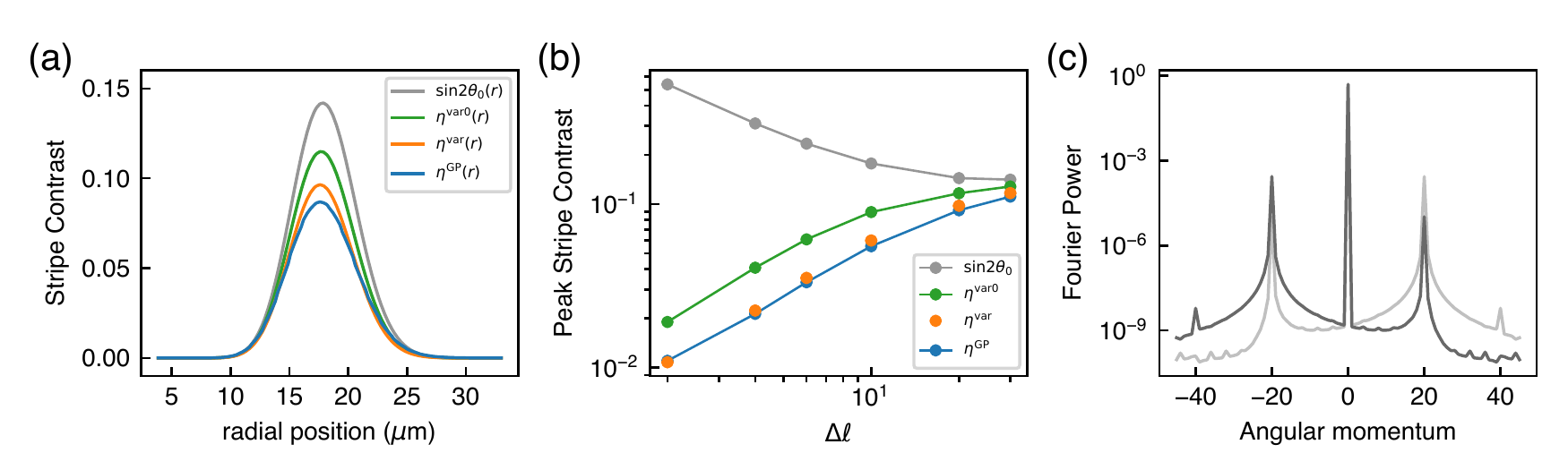}
    \caption{Stripe contrast from the GP and variational
    calculations with varying $\Delta \ell$. $\Omega_M=\Omega_c$, $r_M=17\micron,R_{\rm TF}=46\micron$ and $\mu/h=21$~Hz.
    (a) Contrast versus radial position $r$ for $\Delta \ell=20$. From variational solutions:
    $\sin 2\theta_0(r)$ for non-interacting atoms (grey curve), $\eta^{\rm var0}(r)$ for
    using two-quasiangular-momentum ansatz (green), and $\eta^{\rm var}(r)$ for
    using four-quasiangular-momentum ansatz (orange). Blue curve
    indicates $\eta^{\rm GP}(r)$ for the GP. (b) Peak values of the
    contrast vs. $\Delta \ell$. Grey, green, orange, and blue symbols
    denote $\sin 2\theta_0,\eta^{\rm var0},\eta^{\rm var},\eta^{\rm GP}$, respectively.
    (c) Annular Fourier power spectrum after integration along $r$ for
    $\ket{\downarrow}$ (dark grey) and $\ket{\uparrow}$ (light grey).}
\end{figure*}
By expanding $A_{\pm}^{\rm var}$ to first order in $\bar{n}g_1/E_L$
and $\hbar\Omega/E_L$, we obtain $A_{\pm}^{\rm var} \approx
\bar{n}g_1\theta/\sqrt{2}E_L$, which also agrees with the result
using perturbation (see appendix). After plugging it into
Eq.~\eqref{eq:nc_var_Apm}, we have
\begin{align}\label{eq:theta_var_expand}
\theta^{\rm var} \approx \frac{\hbar\Omega}{2
E_L}(1-\frac{\bar{n}g_1}{E_L}),
\eta^{\rm var} \approx
\frac{\hbar\Omega}{E_L}(1-2\frac{\bar{n}g_1}{E_L}).
\end{align}
Comparing to $\eta^{\rm var0}$ in Eq.~\eqref{eq:theta_var0_expand},
we find the coefficient of $\bar{n}g_1/E_L$ in $\eta^{\rm var}$ is
$-2$, twice as that in $\eta^{\rm var0}$. Including the additional
OAM $\ell=\pm 3\Delta \ell/2$ in Eq.~\eqref{eq:spinor_Psi_Apm} is
necessary for correct results to first order in $\bar{n}g_1/E_L$.
For the separated phase with $C_+=1,C_-=0$ with $A_+= A_-=0$, it is
identical to that using the two-quasiangular-momentum ansatz.

We comment on earlier theoretical papers on SOAMC
systems~\cite{Sun2015,Chen2016,Chen2019}. We examine the peak
dimensionless interaction strength $\bar{n}g_1/E_L$ in these papers.
In Ref.~\cite{Chen2016}, $\bar{n}g_1/E_L$ is $\lesssim 0.01$, and
single-particle eigenstates are taken as the basis of the
variational method, i.e., $\theta=\theta_0$.
Refs.~\cite{Sun2015,Chen2019} use variational methods with the wave
function ansatz Eq.~\eqref{eq:spinor_Psi}, where Ref.~\cite{Sun2015}
has ring traps with $\bar{n}g_1/E_L>100$, and the interaction
$\bar{n}g_1$ is not specified In Ref.~\cite{Chen2019}.

\section{Results and discussions}
We consider practical experimental parameters to maximize the
density contrast of the stripe phase. We first discuss BECs in
harmonic traps in the Thomas-Fermi regime along the radial direction
with the Thomas-Fermi radius $R_{\rm TF}$. We study how the GP
stripe phase contrast depends on ($\Delta \ell,r_M,R_{\rm TF},\mu$);
$\mu$ is the chemical potential and the peak mean field energy in
the harmonic trap.

For comparisons, we also consider atoms in ring traps. Here
$r_M=r_0$ is the only length scale, unlike the harmonically trapped
systems where there are two relevant length scales, ($r_M,R_{\rm
TF}$).

\begin{figure*}
    \centering
    \includegraphics[width=7 in]{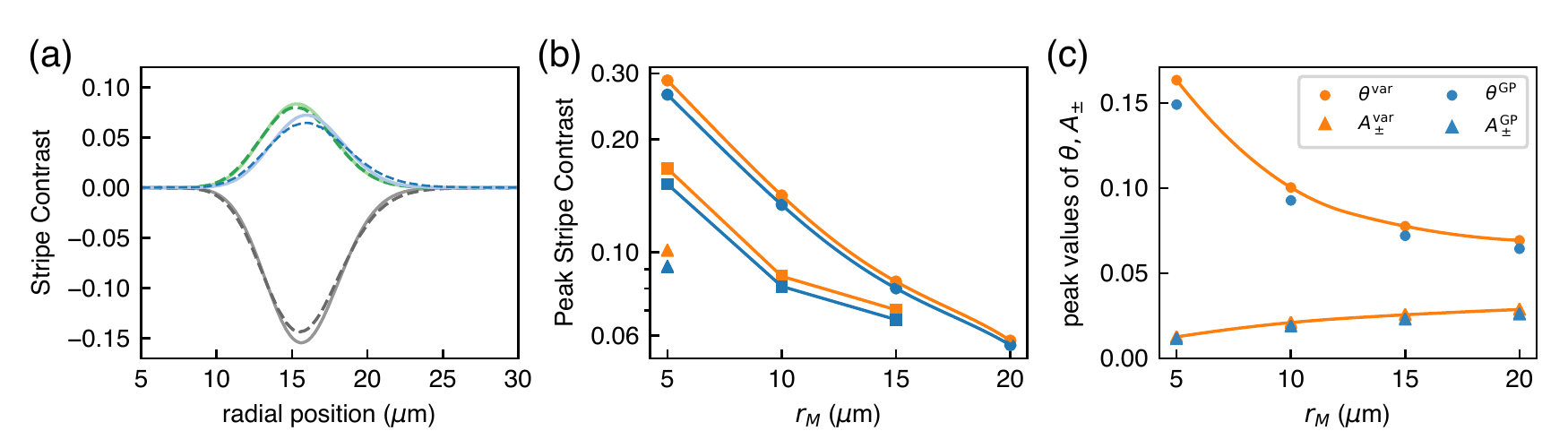}
    \caption{Stripe contrast and $\theta,A_{\pm}$ of the variational results
    using four-quasiangular-momentum
    ansatz and of GP with varying ($r_M,R_{\rm TF}$). $\Omega_M=\Omega_c$,
    $\Delta \ell=20$ and $\mu/h=93$~Hz. (a) Comparison of the
    contrast $\eta$ vs. $r$ for $r_M=15\micron$ and $R_{\rm TF}=50\micron$.
    Grey, blue and green curves indicate $-\sin 2\theta$,
    $2\sqrt{2}\cos\theta A_{\pm}$ and $\eta$, respectively. Solid
    (dashed) curves denote the variational (GP) simulations.
    (b) Peak contrast vs. $r_M$. Orange (blue) symbols
    indicate $\eta^{\rm var} (\eta^{\rm GP})$. Circles, squares
    and triangles for $R_{\rm TF}=50, 25, 12.5\micron$, respectively.
    (c) Peak values of $\theta,A_{\pm}$ vs. $r_M$ for $R_{\rm TF}=50\micron$.
    Orange (blue) symbols indicate the variational (GP) results; circles (triangles)
    for $\theta (A_{\pm})$.}
\end{figure*}

\subsection{Harmonic traps}
We first obtain the GP ground state phase diagram as shown in
Fig.~1c. We then focus on the GP stripe phase at $\delta=0$, setting
$\Omega_M =\Omega_c$. We run simulations for $\Delta \ell$ between 2
and 30, all with $r_M=17~\mu$m, $R_{\rm TF}=46~\mu$m, and
$\mu=h\times 21~$Hz. $\Delta \ell=30$ corresponds to the LG beam
with phase winding number of $\pm 15$, which can be achieved
experimentally (in Ref.~\cite{Tammuz2011}, LG beams with phase
winding number of 45 are realized). From the GP wave function
$\psi(r,\phi)$, we evaluate the density contrast $\eta^{\rm GP}(r)$
from the normalized Fourier components $\tilde{\psi}$ of
$\psi(r,\phi)$ following Eq.~\eqref{eq:contrast_Apm}. For
$\ket{\downarrow}$, $\tilde{\psi}_{\ell_{\downarrow},\downarrow}$
are given by [see Eq.~\eqref{eq:spinor_Psi_Apm}]
\begin{align}
\tilde{\psi}_{0,\downarrow}= -C_+ \cos \theta^{\rm
GP},\tilde{\psi}_{-20,\downarrow}= -C_- \sin \theta^{\rm
GP},\tilde{\psi}_{20,\downarrow}=A_{\pm}^{\rm GP},
\end{align}
from which the contrast $\eta_{\downarrow}^{\rm
GP}=\eta_{\uparrow}^{\rm GP}\approx |\sin 2\theta^{\rm GP} -
2\sqrt{2}\cos \theta^{\rm GP} A_{\pm}^{\rm GP}|$ derived in
Eq.~\eqref{eq:contrast_Apm} can be described as
$4\tilde{\psi}_{0,\downarrow}\tilde{\psi}_{-20,\downarrow}$ (the
first term) and
$4\tilde{\psi}_{0,\downarrow}\tilde{\psi}_{20,\downarrow}$ (the
second term). We then compare $\eta^{\rm GP}(r)$ to the variational
solutions of the contrast, which are $\sin 2\theta_0(r)$ for the
non-interacting case, $\eta^{\rm var0}(r)$ for using the
two-quasiangular-momentum ansatz [Eq.~\eqref{eq:nc_var0} from the
ansatz Eq.~\eqref{eq:spinor_Psi}] and $\eta^{\rm var}(r)$ for using
the four-quasiangular-momentum ansatz [Eq.~\eqref{eq:nc_var_Apm}
from the ansatz Eq.~\eqref{eq:spinor_Psi_Apm}]. In Fig.~3a, we plot
$\sin 2\theta_0(r)$, $\eta^{\rm var0}(r)$, $\eta^{\rm var}(r)$ and
$\eta^{\rm GP}(r)$ for the example value $\Delta \ell=20$; their
maxima are at $r\gtrsim r_M$. In Fig.~3b, We plot the peak values of
$\sin 2\theta_0(r)$, $\eta^{\rm var0}(r)$, $\eta^{\rm var}(r)$ and
$\eta^{\rm GP}(r)$ versus $\Delta \ell$, which are denoted as $\sin
2\theta_0,~\eta^{\rm var0},~\eta^{\rm var}$ and $\eta^{\rm GP}$,
respectively. We observe that the single-particle contrast $\sin
2\theta_0$ is significantly larger than $\eta^{\rm GP}$ for small
$\Delta \ell$, while $\sin 2\theta_0$ and $\eta^{\rm GP}$ are close
for $\Delta \ell \geq 20$. As the dimensionless interaction
$\bar{n}g_1/E_L$ increases with decreasing $\Delta \ell$, the
contrast $\eta^{\rm GP}$ decreases. When the interaction is taken
into consideration using the ansatz Eq.~\eqref{eq:spinor_Psi}, the
resulting $\eta^{\rm var0}$ overestimates $\eta^{\rm GP}$,
indicating that Eq.~\eqref{eq:spinor_Psi} is insufficient. We can
understand this from the annular Fourier transform of the GP wave
function $\psi_{\downarrow}$ for $\Delta \ell=20$. Fig.~3c shows the
power spectrum of the normalized Fourier components
$\tilde{\psi}_{\ell_{\downarrow},\downarrow}$, where there are
$\ell_{\downarrow}=0, \pm20$ components, and the $\ell_{\downarrow}=
-40$ is negligible since its power is about $10^{-3}$ of that of
$\ell_{\downarrow}=20$. The appearance of the $\ell_{\downarrow}=20$
component signifies that the more general
Eq.~\eqref{eq:spinor_Psi_Apm} should be used in the variation method
with the $A_+$ term accounting for $\ell_{\downarrow}=20$, while
$A_+,A_-$ are absent in the simple Eq.~\eqref{eq:spinor_Psi}. The
spectrum for $\ket{\uparrow}$ is also displayed in Fig.~3c. The GP
results have $C_{\pm}\approx \mp 1/\sqrt{2}$ and
$A_+=A_-=A_{\pm}^{\rm GP}$, which confirms the time-reversal
symmetry condition, Eq.~\eqref{eq:Cpm_Apm}. [From the GP results,
the signs of $C_{\pm}\approx \mp 1/\sqrt{2}$ are applied in
Eq.~\eqref{eq:Cpm_Apm} and in the variational method using
Eq.~\eqref{eq:spinor_Psi_Apm}.] We also show the peak values,
$\eta^{\rm var}$ of $\eta^{\rm var}(r)$, in Fig.~3b, where
$\eta^{\rm var0}>\eta^{\rm var}\gtrsim \eta^{\rm GP}$ and $\eta^{\rm
var}$ fits well with $\eta^{\rm GP}$.

From the above studies, we find the maximum of the stripe contrast
is at $r=r_{\rm peak}\approx r_M$, where the spatial period is
$\approx 2\pi r_M/ \Delta \ell$. ($r_{\rm peak}$ is only slightly
larger than $r_M$ for all the contrasts, e.g., $r_{\rm
peak}=17.6\micron $ for $\eta^{\rm GP}$ in Fig.~3a.) The peak value
of the contrast increases with $\Delta \ell$, and thus a larger
contrast corresponds to a small stripe period $\propto \Delta
\ell^{-1}$. To observe the stripe phase in experiments, we note that
with state-of-the-art imaging techniques in ultracold atoms, e.g.
those using quantum gas microscopes, one can resolve as small as
$0.5\micron$ with $\lambda= 0.78\micron$ for
$^{87}$Rb~\cite{Bakr2009}. This sets the lower bound on $2\pi r_M/
\Delta \ell$ in our simulations. Since the peak contrast is the
signal we optimize, $r_{\rm peak}\approx r_M$ and thus $E_L(r_{\rm
peak})\approx E_L(r_M)$ are the relevant length and energy scale for
SOAMC, respectively.

Next, we fix $\Delta \ell= 20$ and $\mu=h\times 93~$Hz, and vary
$(r_M,R_{\rm TF})$. We study $r_M<R_{\rm TF}$ where there is
sufficient atomic density at $r=r_M$ for various combinations of
$(r_M,R_{\rm TF})$. Here we set the smallest $r_M=5~\mu$m for
$\Delta \ell=20$, where the spatial period of the stripe is $\approx
1.6\micron$ and is larger than the diffraction limit of the imaging,
$0.5\micron$. With $\Omega_M=\Omega_c$, the GP results have the peak
contrast $\eta^{\rm GP}$ increasing with decreasing $r_M$ and
increasing $R_{\rm TF}$, as shown in Fig.~4b. We then compare the GP
to the variational calculations with four-quasiangular-momentum
ansatz . In Fig.~4a, we plot the density contrast contributed from
$\theta,A_{\pm}$ and the sum, respectively. These are $-\sin
2\theta^{\rm GP}$, $2\sqrt{2}\cos \theta^{\rm GP} A_{\pm}^{\rm GP}$,
and $\eta^{\rm GP}$ versus $r$ for GP, and $-\sin 2\theta^{\rm
var}$, $2\sqrt{2}\cos \theta^{\rm var} A_{\pm}^{\rm var}$, and
$\eta^{\rm var}$ versus $r$ for the variational calculations. We
find the contrast of GP obtained from Eq.~\eqref{eq:nc_max_min}
versus $r$ agrees well with $\eta^{\rm GP}(r)$, showing that
$\eta^{\rm GP}$ from the $\cos (\Delta \ell\phi)$ term dominates the
contrast in Eq.~\eqref{eq:density_stripe_Apm}, where the second
harmonics is negligible. We display the peak values $\eta^{\rm var}$
versus $r_M$ for all $R_{\rm TF}$ in Fig.~4b and compare them to the
peak values $\eta^{\rm GP}$, where $\eta^{\rm var}$ overestimates
$\eta^{\rm GP}$ by $3-8~\%$. For all ($r_M,R_{\rm TF}$), the Fourier
spectrum of GP has $\ell_{\uparrow,\downarrow}=0, \pm20$ components
and $\ell_{\uparrow}=40,\ell_{\downarrow}= -40$ are negligible,
being consistent with Eq.~\eqref{eq:spinor_Psi_Apm}. We then compare
the peak values $(\theta^{\rm GP},A_{\pm}^{\rm GP})$ to
$(\theta^{\rm var},A_{\pm}^{\rm var})$ versus $r_M$ for $R_{\rm
TF}=50~\mu$m in Fig.~4c. The peaks of $\theta$ and $A_{\pm}$ are at
$r\approx r_M$, as well as that of the contrast $\eta$. $\theta^{\rm
var}$ and $A_{\pm}^{\rm var}$ slightly overestimate $\theta^{\rm
GP}$ and $A_{\pm}^{\rm GP}$, respectively: both $\theta^{\rm
GP}/\theta^{\rm var}$ and $A_{\pm}^{\rm GP}/A_{\pm}^{\rm var}$ are
between $0.90-0.93$. This is attributed to the radial kinetic energy
that is neglected in the variational calculations. The GP ground
state has smaller $(\theta^{\rm GP},A_{\pm}^{\rm GP})$ than
$(\theta^{\rm var},A_{\pm}^{\rm var})$ where the smaller radial spin
gradient corresponds to a smaller radial kinetic energy, and thus
the lowest overall energy.

\begin{figure}
    \centering
    \includegraphics[width=3.0 in]{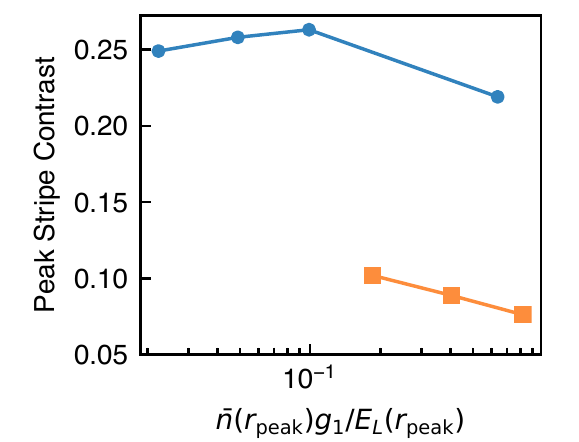}
    \caption{Peak stripe contrast $\eta^{\rm GP}$ versus interaction strength
    $\bar{n}(r_{\rm peak})g_1/E_L(r_{\rm peak})$ with varying $\mu$, $\Delta \ell=20$ and $R_{\rm TF}=50~\mu$m.
    Circles and square symbols indicate $r_M=5,15~\mu$m, respectively.}
\end{figure}

After studying the dependence of the peak contrast $\eta^{\rm GP}$
on ($\Delta \ell,r_M,R_{\rm TF}$), we vary $\mu$ and thus the
interaction strength $\bar{n}g_1/E_L$ at $r=r_{\rm peak}\approx r_M$
, $\bar{n}g_1/E_L=\mu \left[1-(r_{\rm peak}/R_{\rm
TF})^2\right]/E_L(r_{\rm peak})$. We fix $\Delta \ell=20, R_{\rm
TF}=50\micron$ for $r_M=5$ and $15\micron$, respectively. We study
$\mu=h \times 21,46, 93$~Hz for both $r_M$, and additionally $\mu=h
\times 600$~Hz for $r_M=5\micron$. Fig.~5 shows $\eta^{\rm GP}$
weakly depends on $\bar{n}g_1/E_L$.

Besides the density contrast, we compare the critical coupling
$\Omega_c$ of the GP results to those given by the variational
methods. Using the two-quasiangular-momentum ansatz,
Eq.~\eqref{eq:spinor_Psi}, the critical coupling $\Omega_c^{\rm
var0}$ is given by
\begin{align}\label{eq:Omegac_var0}
&\int dr 2\pi r \bar{n}(r) \Delta\varepsilon^{\rm var0}=0,\nonumber\\
&\Delta\varepsilon^{\rm var0}=\varepsilon^{\rm var0}(\theta^{\rm
var0},\beta=1/4)-\varepsilon_{\rm sep}.
\end{align}
$\Delta\varepsilon^{\rm var0}$ is the energy difference between the
stripe phase and the separated phase, $\theta^{\rm var0}$ is the
variational solution of $\theta$ for the stripe phase, and
$\varepsilon_{\rm sep}$ is the energy of the separated phase with
$\theta=\theta_{\rm sep}$ and $\beta=0$. When the integral is
$<(>)~0$ at $\Omega_M< (>) \Omega_c$, the ground state is the stripe
(separated) phase. Similarly, by using the
four-quasiangular-momentum ansatz, Eq.~\eqref{eq:spinor_Psi_Apm},
the critical coupling $\Omega_c^{\rm var}$ is given by

\begin{align}\label{eq:Omegac_var_Apm}
&\int dr 2\pi r \bar{n}(r)\Delta\varepsilon^{\rm var} =0,\nonumber\\
&\Delta\varepsilon^{\rm var}=\varepsilon^{\rm var}(\theta^{\rm
var},C_{\pm}=\mp \sqrt{\left[1-2(A^{\rm var}_{\pm})^2
\right]/2})-\varepsilon_{\rm sep}.
\end{align}

\begin{figure}
    \centering
    \includegraphics[width=3.0 in]{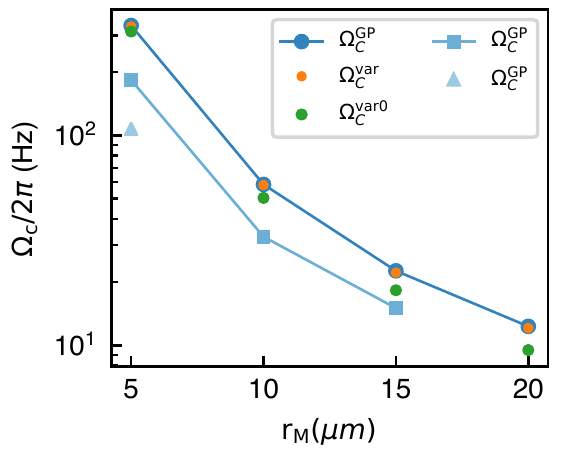}
    \caption{Critical coupling $\Omega_c$ vs. $r_M$ with various $R_{\rm TF}$ of GP
    and variational results from two-quasiangular-momentum and
    four-quasiangular-momentum ansatz. Blue symbols denote $\Omega_c^{\rm GP}$; circles, squares
     and triangle for $R_{\rm TF}=50,25,12.5\micron$, respectively. Orange (green) symbols indicate
    $\Omega_c^{\rm var}(\Omega_c^{\rm var0})$ for $R_{\rm TF}=50\micron$.}
\end{figure}

In Fig.~6, we plot $\Omega_c$ of the GP and from the solutions of
Eq.~\eqref{eq:Omegac_var0} and Eq.~\eqref{eq:Omegac_var_Apm} versus
$(r_M,R_{\rm TF})$, where $\Omega_c$ from GP have good agreements
with $\Omega_c^{\rm var}$.


\begin{figure*}
    \centering
    \includegraphics[width=6.0 in]{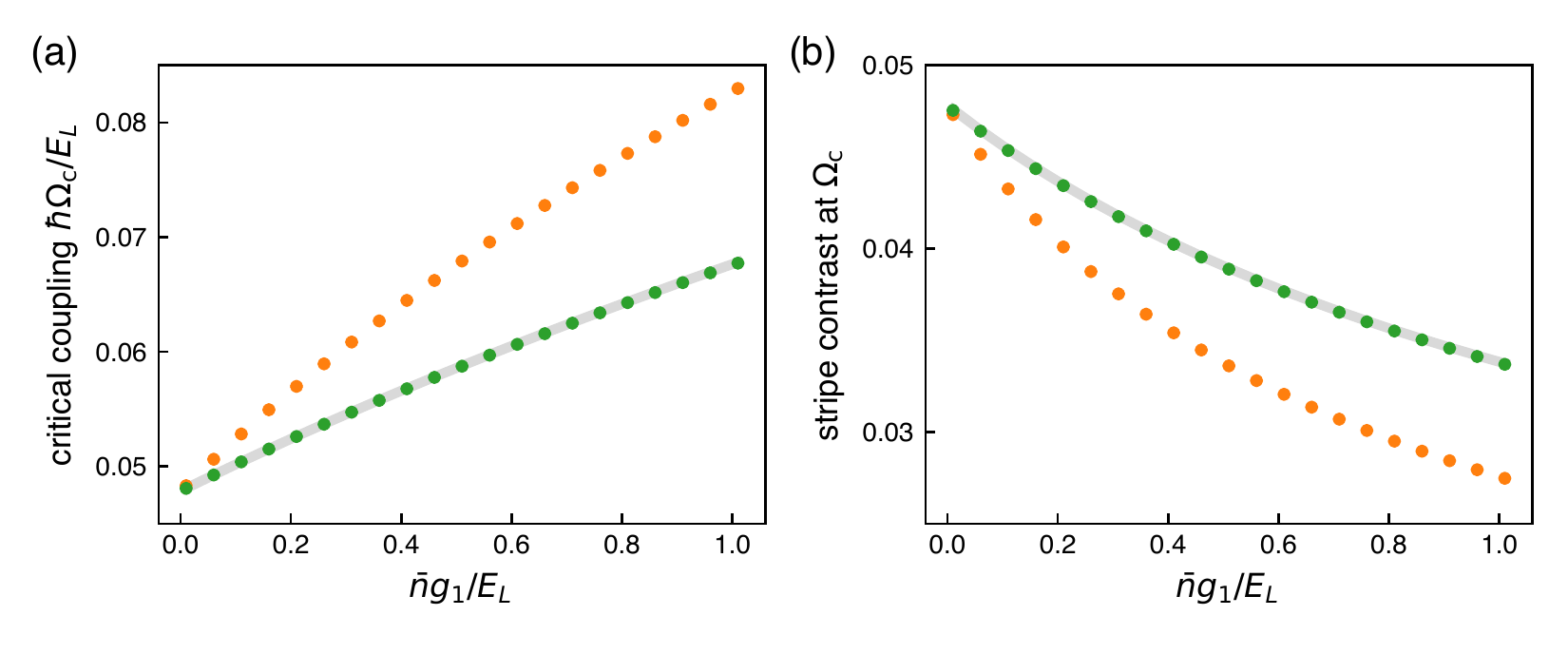}
    \caption{Critical coupling $\hbar\Omega_c/E_L$ and stripe contrast
    $\eta$ at $\Omega_c$ vs. $\bar{n}g_1/E_L$ at fixed $r$. Variational
    calculations of two-quasiangular-momentum and
    four-quasiangular-momentum ansatz for SOAMC, and of
    two-quasilinear-momentum ansatz for SLMC are illustrated.
    Green symbols, orange symbols and grey curve display
    $\hbar\Omega_c^{\rm var0}/E_L,\hbar\Omega_c^{\rm var}/E_L$
    and $\hbar\Omega_c^{\rm SLMC}/E_L$, respectively for (a), and display $\eta^{\rm var0},\eta^{\rm var}$
    and $\eta^{\rm SLMC}$ for (b).}
\end{figure*}

We can understand that $\Omega_c$ increases with increasing $R_{\rm
TF}$ and decreasing $r_M$ from a geometric argument. Such dependence
on ($r_M,R_{\rm TF}$) is crucial since a larger $\hbar\Omega_c/E_L$
leads to larger stripe contrast $\eta^{\rm var}$, see
Eq.~\eqref{eq:theta_var_expand}. We find numerically that the energy
difference between the stripe phase and separated phase can be
written as
\begin{align}\label{eq:Omegac_var_Apm2}
\Delta\varepsilon^{\rm var}&=f(r)-\frac{\bar{n}(r)g_2}{2},\nonumber\\
f(r)&=\bar{n}(r)g_1
H(\frac{\bar{n}g_1}{E_L})\frac{\hbar^2\Omega^2(r)}{E_L^2}
\end{align}
for small $\theta$, where $H$ is a dimensionless function of
$\bar{n}g_1/E_L$ and $H\rightarrow 1/4$ as
$\bar{n}g_1/E_L\rightarrow 0$. To make a geometric analysis, we
simplify $f(r)$ as a flat impulse function centered at $r=r_M$ with
full width $\Delta r=(\Delta \ell /2)^{-1/2}r_M$,
\begin{align}\label{eq:f_M}
f(r)&=f_M, r_M-\Delta r/2 < r < r_M+ \Delta r/2,\nonumber\\
f_M&=\bar{n}g_1
H(\frac{\bar{n}g_1}{E_L})\frac{\hbar^2\Omega_M^2}{E_L^2},
\end{align}
where $\bar{n}g_1,E_L$ and $\Omega_M$ are evaluated at $r=r_M$,
approximately the peak position of $f(r)$. Assuming a cylindrical
box trap with uniform $\bar{n}$ within $r=R_{\rm box}$, the integral
in Eq.~\eqref{eq:Omegac_var_Apm} then gives
\begin{align*}
\bar{n} 2\pi r_M \Delta r f_M=\bar{n}\pi R_{\rm box}^2
\frac{\bar{n}g_2}{2},
\end{align*}
and thus
\begin{align}\label{eq:f_M2}
f_M=\frac{\bar{n}g_2}{2}\left( \frac{\pi R_{\rm box}^2}{2\pi r_M
\Delta r}\right),
\end{align}
where the number in the parentheses is an area ratio of the box to
that of the distribution $f(r)$. From Eq.~\eqref{eq:f_M} and
Eq.~\eqref{eq:f_M2} and assuming a fixed interaction strength
$\bar{n}g_1/E_L$, we obtain $\Omega_M=\Omega_c \propto R_{\rm box}
r_M^{-3}$ which increases with increasing $R_{\rm box}$ and
decreasing $r_M$.

\subsection{Ring traps}
We show the variational results for the ring trap versus the
dimensionless interaction strength
$g_1^{\prime}=\bar{n}g_1/E_L(r_0)$. These are the solutions of the
critical coupling $\Omega_c$ and the contrast at $\Omega_c$. We
solve the dimensionless critical coupling
$\Omega_c^{\prime}=\hbar\Omega_c/E_L$ by using
$\Delta\varepsilon^{\rm var0}(r_0)=0$ and $\Delta\varepsilon^{\rm
var}(r_0)=0$, respectively, and plot $\Omega_c^{\prime}$ vs.
$g_1^{\prime}$ in Fig.~7a. $\Omega_c^{\rm var}$ exceeds
$\Omega_c^{\rm var0}$ for $g_1^{\prime}>0$, and $\hbar\Omega_c^{\rm
var0} \approx \hbar\Omega_c^{\rm var} \rightarrow \sqrt{2
g_2/g_1}~E_L\approx 0.05~E_L$ as $g_1^{\prime}\rightarrow 0$. we can
understand $\Omega_c^{\rm var}>\Omega_c^{\rm var0}$ as the
following: At a given $\Omega$, the stripe phase energy is
$\varepsilon^{\rm var}<\varepsilon^{\rm var0}$ since the smaller
contrast $\eta^{\rm var}$ [see Eq.~\eqref{eq:theta_var0_expand} and
Eq.~\eqref{eq:theta_var_expand}] corresponds to smaller interaction
energy. Therefore, the critical coupling determined by $\Delta
\varepsilon=0$, where $\Delta \varepsilon$ increases with $\Omega$,
is shifted to a larger value for $\Omega_c^{\rm var}$. In the
$g_1^{\prime}\rightarrow 0$ limit, $\Delta\varepsilon^{\rm
var0}(r_0)\approx \Delta\varepsilon^{\rm var}(r_0)$, where the
stripe and separated phases have approximately the same variational
solution, $\theta^{\rm var0} \approx \theta_{\rm sep} \approx
\theta_0$, and thus $\Delta\varepsilon^{\rm var0}=(1/4)
\bar{n}\left[g_1\sin^2 2\theta_0-2 g_2\cos^2 2\theta_0 \right]=0$
from Eq.~\eqref{eq:E_variational}, leading to
$\Omega_c^{\prime}=\sqrt{2 g_2/g_1}$. The stripe contrasts
$\eta^{\rm var0}$ and $\eta^{\rm var}$ at respective $\Omega_c$ are
displayed in Fig.~7b; they are $\approx 5~\%$ as
$g_1^{\prime}\rightarrow 0$, both decreasing with increasing
$g_1^{\prime}$ and $\eta^{\rm var}<\eta^{\rm var0}$. To compare with
SLMC systems, $\hbar\Omega_c^{\rm var0}/E_L$ for $g_1^{\prime}<1$
agrees well with that of SLMC (see Fig.~7a), which is
$\hbar\Omega_c^{\rm SLMC}/4E_r=\sqrt{2 g_2/g_1}$ as
$g_1^{\prime}\rightarrow 0$~\cite{Li2012}, i.e., $\hbar\Omega_c^{\rm
SLMC}= 0.2 E_r$; $4 E_r$ is equivalent to $E_L$ for SOAMC.

The spin-dependent interaction strength $g_2/g_1$ determines the
critical coupling $\hbar \Omega_c/E_L$ and stripe contrast $\eta$ as
$g_1^{\prime}\rightarrow 0$; larger $g_2/g_1$ gives larger $\hbar
\Omega_c/E_L$ and $\eta$, i.e., larger miscibility. The contrast of
SLMC also agrees well with $\eta^{\rm var0}$, see Fig.~7b. We find
that $\Omega_c^{\prime}$ and $\eta$ of SOAMC with $\varepsilon^{\rm
var0}$ and of SLMC, where both are based on
Eq.~\eqref{eq:spinor_Psi}, are incorrect to first order in
$g_1^{\prime}$. By including higher order OAM in
Eq.~\eqref{eq:spinor_Psi_Apm}, the result is correct to first order,
as indicated by Eq.~\eqref{eq:theta_var0_expand} and
Eq.~\eqref{eq:theta_var_expand}. The stripe contrast of SOAMC is at
most $\approx 5~\%$ in the $g_1^{\prime}\rightarrow 0$ limit, and it
is independent of either the ring radius $r_0$ or $\Delta \ell$. On
the other hand, a stripe contrast $\lesssim 30~\%$ for harmonic
traps is achieved with a relatively large $R_{\rm TF}=50\micron$ and
a relatively small $r_M=5\micron$, and this can be understood from
the geometric analysis as shown in Eq.~\eqref{eq:f_M2}.

We then perform GP simulations for a ring trap with
$r_0=r_M=10\micron$ and $\Delta \ell=20$. The atoms are in an
annular box potential within $8\micron< r < 12\micron$, and
$\bar{n}g_1/E_L=0.63$ at $r=r_0$. The GP result has good agreement
with the variational calculation, where $\theta^{\rm GP}/\theta^{\rm
var}=0.95$ and $A_{\pm}^{\rm GP}/A_{\pm}^{\rm var}=0.91$.

\section{Conclusions}
In summary, we optimize the density contrast of the ground state
stripe phase of $^{87}$Rb SOAMC BECs by tuning experimental
parameters. A contrast of nearly $30~\%$ is achieved for atoms in
harmonic traps; and a larger contrast of about $50~\%$ is expected
by using a twice larger BEC cloud size based on variational
calculations. Such high contrasts are achieved owing to the geometry
with two length scales in harmonic traps, the Raman
Laguerre-Gaussian beam size and the BEC cloud size. While for ring
traps, these two scales are the same, leading to maximal contrast
about $5~\%$, which is dictated by the spin-dependent interaction
strength and is the same as that of the SLMC systems. For both atoms
in harmonic traps and ring traps, we perform GP simulations and
variational calculations based on the two-quasiangular-momentum
ansatz and four-quasiangular-momentum ansatz. We find the results
from the simple two-quasiangular-momentum ansatz, which is used in
previous papers~\cite{Li2012,Sun2015,Chen2016,Chen2019}, is
consistent with the GP results only in the non-interacting limit.
With small interactions, high order OAM components must be included
as the four-quasiangular-momentum ansatz; this then leads to correct
results to first order in interaction and good agreements with the
GP simulations.

We point out that one can improve the stability by using the
synthetic clock states instead of bare spin states. The clock states
are immune to detuning variations arising from the bias field
variations. A $0.1-1$ Hz stability of the clock transition frequency
is achieved as shown in Ref.~\cite{Trypogeorgos2018}. Thus, the
ground state stripe phase within a narrow detuning window of about
$1$~Hz may be observed for $^{87}$Rb atoms with mean field energy
about 1~kHz. The spin-dependent interaction strength in the clock
state basis is close to the $g_2/g_1$ for bare spin states (see
appendix), leading to similar magnitude of stripe contrast to our
simulations using bares spin states. We envision our work to pave
the way toward a direct observation of high-contrast stripe phases
in spin-orbital-angular-momentum coupled Bose-Einstein condensates,
achieving a long-standing goal in quantum gases.

\section{Appendix}
\subsection{Spinor wave function ansatz}
We show that the spinor wave function ansatz
Eq.~\eqref{eq:spinor_Psi_Apm} is valid for small $\theta$ (given
small Raman coupling $\hbar\Omega/E_L$) and small interaction
$\bar{n}g_1/E_L$. The GPE for $\ket{\uparrow},\ket{\downarrow}$ with
$\delta=0$ is
\begin{subequations}
\begin{align}\label{eq:GPE_updown}
&(-\frac{\hbar^2}{2m}\nabla^2
+g|\psi_{\uparrow}|^2+g_{\uparrow\downarrow}|\psi_{\downarrow}|^2 -\mu_{\uparrow} )\psi_{\uparrow}+\frac{\hbar\Omega(r)}{2}e^{i \Delta \ell
\phi}\psi_{\downarrow}=0,\\
&(-\frac{\hbar^2}{2m}\nabla^2
+g|\psi_{\downarrow}|^2+g_{\uparrow\downarrow}|\psi_{\uparrow}|^2-\mu_{\downarrow}) \psi_{\downarrow}+\frac{\hbar\Omega(r)}{2}e^{-i \Delta \ell
\phi}\psi_{\uparrow}
=0.
\end{align}
\end{subequations}
Here we set $V(r)=0$ to simplify the discussion. For the
spatially-mixed stripe phase ground state,
$\mu_{\uparrow}=\mu_{\downarrow}$. Next we show the nonlinear
interaction leads to multiple OAM components in the spinor wave
function ansatz. With $\ell=\Delta \ell/2=10$, we plug
$\psi_{\uparrow}=\sqrt{\bar{n}(r)}\left(C_+\sin \theta e^{i
20\phi}+C_-\cos \theta
\right),\psi_{\downarrow}=\sqrt{\bar{n}(r)}\left(-C_+\cos \theta
-C_-\sin \theta e^{-i 20\phi} \right)$ from
Eq.~\eqref{eq:spinor_Psi} into Eq.~\eqref{eq:GPE_updown}, neglect
radial gradients and keep the expansion terms up to $\theta^2$. The
nonlinear interaction terms for $\psi_{\uparrow}$ are

\begin{align}\label{eq:nonlinear}
g|\psi_{\uparrow}|^2\psi_{\uparrow}+g_{\uparrow\downarrow}|\psi_{\downarrow}|^2\psi_{\uparrow}&=\sqrt{\bar{n}}^3[ C_-C_+^2(g+g_{\uparrow\downarrow})\theta^2e^{i 40\phi} \nonumber \\
&+C_+\left(2C_-^2g+g_{\uparrow\downarrow}\right)\theta e^{i 20\phi}\nonumber\\
&+C_-\left(C_-^2 g+C_+^2 g_{\uparrow\downarrow}+O[\theta^2]\right) \nonumber\\
&+C_-^2C_ +\left(g+g_{\uparrow\downarrow}\right)\theta e^{-i20\phi}].
\end{align}

Besides $\ell_{\uparrow}=0,20$, additional OAM terms with
$\ell_\uparrow=-20, 40$ appear due to the nonlinear interaction,
which are of order of $\theta,\theta^2$, respectively, and are not
included in Eq.~\eqref{eq:spinor_Psi}. By keeping up to order
$\theta^2$, the spinor wave function has additional variational
parameters $A_+,A_-,B_+,B_-$, given by

\begin{align}\label{eq:spinor_Psi_Bpm}
\left(\begin{array}{c} \psi_{\uparrow}\\
\psi_{\downarrow}
\end{array}\right) = \sqrt{\bar{n}(r)}[&e^{i \frac{3\Delta \ell}{2}\phi}\left(
\begin{array}{c}
B_+ e^{i\frac{\Delta \ell}{2}\phi}\\
A_+ e^{-i\frac{\Delta \ell}{2}\phi}
\end{array}\right) \nonumber \\ 
&+C_+ e^{i \frac{\Delta \ell}{2}\phi}\left(
\begin{array}{c}
\sin \theta(r)e^{i\frac{\Delta \ell}{2}\phi}\\
-\cos \theta(r)e^{-i\frac{\Delta \ell}{2}\phi}
\end{array}\right) \nonumber\\
&+ C_- e^{-i\frac{\Delta \ell}{2}\phi}\left(
\begin{array}{c}
\cos \theta(r)e^{i\frac{\Delta \ell}{2}\phi} \\
-\sin \theta(r)e^{-i\frac{\Delta \ell}{2}\phi}
\end{array}\right) \nonumber \\
&+ e^{-i\frac{3\Delta \ell}{2}\phi}\left(
\begin{array}{c}
A_- e^{i\frac{\Delta \ell}{2}\phi}\\
B_- e^{-i\frac{\Delta \ell}{2}\phi}
\end{array}\right)].
\end{align}
$\ell_\uparrow= -20$ is of order $\theta$ and corresponds to
$\ell=-3\Delta \ell/2=-30$ with $A_-$; $\ell_\uparrow= 40$ is of
order $\theta^2$ and corresponds to $\ell=3\Delta \ell/2=30$ with
$B_+$. For small $\theta$, by taking up to order $\theta$ we have
the spinor wave function ansatz Eq.~\eqref{eq:spinor_Psi_Apm} with
$A_+,A_-\neq 0$ and $B_+=B_-=0$.

\begin{figure*}
    \centering
    \includegraphics[width=6.0 in]{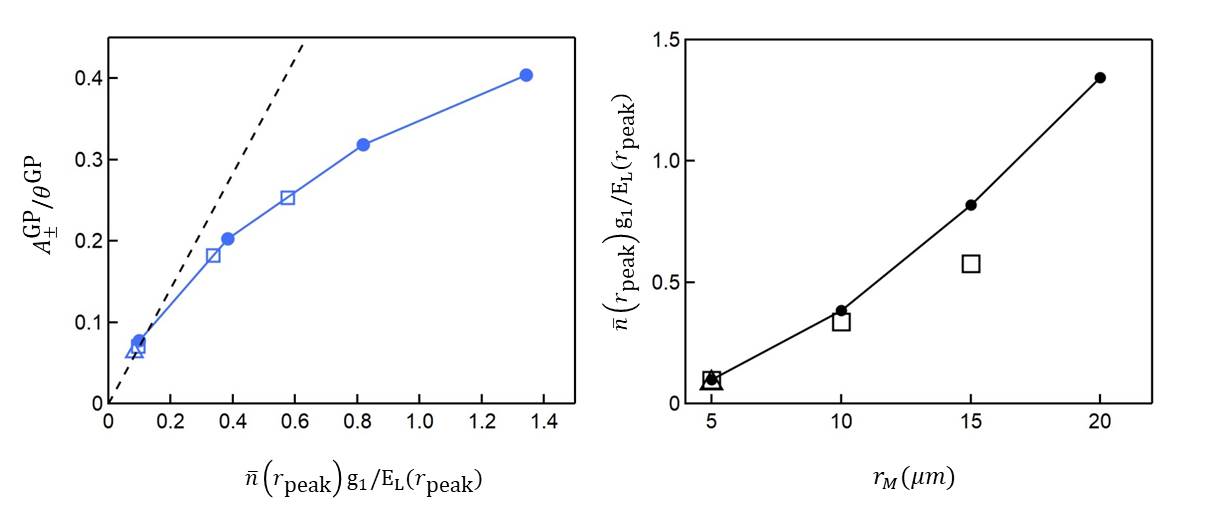}
    \caption{(a) Ratio of the peak values $A_{\pm}^{\rm GP}/\theta^{\rm GP}$ of GP stripe phase
    versus the interaction evaluated at $r_{\rm peak}$,  $\bar{n}(r_{\rm peak})g_1/E_L(r_{\rm peak})$.
    Circles for $R_{\rm TF}=50\micron$, open squares for $25\micron$, and open triangle for $12.5\micron$;
    $r_{\rm peak}\approx r_M$. (b) $\bar{n}(r_{\rm peak})g_1/E_L(r_{\rm peak})$ versus $r_M$
    for $R_{\rm TF}=50\micron$, $25\micron$, and $12.5\micron$. Same symbols as (a) and all with
    $\mu/h=93$~Hz.}
\end{figure*}

Next we derive $A_-$ for small interactions $\bar{n}g_1/E_L$ using
first order perturbation. We plug
\begin{align}
\psi_{\uparrow}&=\sqrt{\bar{n}}\left(C_+\sin \theta e^{i
20\phi}+C_-\cos \theta+A_-e^{-i 20\phi}
\right),\nonumber\\
\psi_{\downarrow}&=\sqrt{\bar{n}}\left(A_+e^{i 20\phi}-C_+\cos
\theta -C_-\sin \theta e^{-i 20\phi} \right)
\end{align}
into Eq.~\eqref{eq:GPE_updown} for $\psi_{\uparrow}$, and focus on
the coefficient of the $e^{-i 20\phi}$ term, which is
\begin{align}
\left(E_L -\mu_{\uparrow}\right)\sqrt{\bar{n}}A_-
+\sqrt{\bar{n}}^3C_-^2C_
+\left(g+g_{\uparrow\downarrow}\right)\theta = 0.
\end{align}
Besides reading out from Eq.~\eqref{eq:nonlinear}, the coefficient
of the $e^{-i 20\phi}$ term in the nonlinear interaction
$g|\psi_{\uparrow}|^2\psi_{\uparrow}+g_{\uparrow\downarrow}|\psi_{\downarrow}|^2\psi_{\uparrow}$
can be readily found from the Fourier components of
$|\psi_{\uparrow}|^2,|\psi_{\downarrow}|^2$ in
Eq.~\eqref{eq:density_stripe}, both of which have OAM$=0,\pm 20$.
Then, using $\mu_{\uparrow} \approx
\mu_{\uparrow}(\Omega=0,g=g_{\uparrow\downarrow}=0)=0$ for small
$\Omega/E_L$ and $\bar{n}g_1/E_L$, along with $C_{\pm}\approx \mp
1/\sqrt{2}$, it gives
\begin{align}\label{eq:Am_perturb}
A_- \approx \frac{\bar{n}g_1}{\sqrt{2}E_L}\theta.
\end{align}
From the GP stripe phase wave function, in Fig.~8a we plot the ratio
of the peak values $A_{\pm}^{\rm GP}/\theta^{\rm GP}$ vs.
$\bar{n}(r_{\rm peak})g_1/E_L(r_{\rm peak})$, along with the ratio
$\bar{n}g_1/\sqrt{2}E_L$ given by Eq.~\eqref{eq:Am_perturb}, which
agrees with $A_{\pm}^{\rm GP}/\theta^{\rm GP}$ at small
$\bar{n}(r_{\rm peak})g_1/E_L(r_{\rm peak})$. The dimensionless
interaction $\bar{n}g_1/E_L$ evaluated at $r_{\rm peak}\approx r_M$
vs. $r_M$ for $R_{\rm TF}=12.5, 25, 50\micron$ is shown in Fig.~8b,
all with $\mu=h \times 93$~Hz. For a fixed $\mu$,
$\bar{n}g_1=\mu[1-(r/R_{\rm TF})^2]$ weakly depends on $r$ for
$r/R_{\rm TF}\lesssim 0.5$. $\bar{n}(r_{\rm peak})g_1/E_L(r_{\rm
peak})$ increases with increasing $r_M$, which is dominated by
$E_L\propto r^{-2}$.

Similarly, we derive $B_+$ by plugging
\begin{align}
\psi_{\uparrow}&=\sqrt{\bar{n}}\left(B_+ e^{i 40\phi}+C_+\sin \theta
e^{i 20\phi}+C_-\cos \theta+A_-e^{-i 20\phi}
\right),\nonumber\\
\psi_{\downarrow}&=\sqrt{\bar{n}}\left(A_+e^{i 20\phi}-C_+\cos
\theta -C_-\sin \theta e^{-i 20\phi} +B_- e^{-i 40\phi} \right)
\end{align}
into Eq.~\eqref{eq:GPE_updown}. The coefficient of the $e^{i
40\phi}$ term is
\begin{align}
\left(4E_L -\mu_{\uparrow}\right)\sqrt{\bar{n}}B_+ +\frac{\hbar
\Omega}{2} \sqrt{\bar{n}} A_+ +\sqrt{\bar{n}}^3 C_-
C_+^2\left(g+g_{\uparrow\downarrow}\right)\theta^2 = 0.
\end{align}
With $\theta \approx \hbar \Omega/2E_L (1-\bar{n}g_1/E_L)$ from
$\theta^{\rm var}$, and $A_+=A_-$ for the ground state, it leads to
\begin{align}
B_+\approx -\left(\frac{\hbar
\Omega}{8E_L}\frac{\bar{n}g_1}{\sqrt{2}E_L}\theta+\frac{\bar{n}g_1}{4\sqrt{2}E_L}\theta^2\right)
\approx -\frac{\bar{n}g_1}{2\sqrt{2}E_L}\theta^2.
\end{align}

In our GP data, the peak values of $B_{\pm}^{\rm GP}$ at $r\gtrsim
r_M$ are small, $0.02 < B_{\pm}^{\rm GP}/A_{\pm}^{\rm GP}< 0.04$ for
$\bar{n}(r_{\rm peak})g_1/E_L(r_{\rm peak})<2$. Thus it is valid to
neglect $B_{\pm}$ by using the wave function ansatz
Eq.~\eqref{eq:spinor_Psi_Apm}. We have small $\theta^{\rm GP}$,
$0.05<\theta^{\rm GP}<0.16$, and small interaction, $\bar{n}(r_{\rm
peak})g_1/E_L(r_{\rm peak})<2$ except for the data with the smallest
$\Delta \ell=2, 4$ in Fig.~3.

\subsection{Methods for GP ground state simulations}
We run the GP simulations with both the open-source GPELab
toolbox~\cite{Antoine2014} and Crank-Nicolson method. The grid size
is between $0.11-0.55\micron$ depending on the spatial resolution we
need.

To do analysis of the GP wave function in the cylindrical
coordinate, we first make interpolations of the raw data in the
cartesian coordinate. The annular Fourier transform is performed as
\begin{align}
\psi_{m}(r,\phi)= \sum_{q^{\prime}} \psi_{q^{\prime},m}(r)
e^{iq^{\prime}\phi},\nonumber\\
\psi_{q,m}=(2\pi)^{-1}\int d\phi \psi_{m}(r,\phi)e^{-iq\phi},
\end{align}
where $q=\ell_{\uparrow},\ell_{\downarrow}$ is the OAM and
$m=\uparrow,\downarrow$ is the spin label. For the stripe phase with
$\bar{n}_{\uparrow}(r)=\bar{n}_{\downarrow}(r)=\bar{n}(r)/2$,
\begin{align}
\sum_{q}|\psi_{q,m}(r)|^2=\bar{n}(r)/2.
\end{align}
We take the normalized Fourier components as
$\tilde{\psi}_{q,m}(r)=\psi_{q,m}(r) \bar{n}(r)^{-1/2}$, leading to
$\tilde{\psi}_{0,\downarrow}= -C_+ \cos \theta^{\rm
GP},\tilde{\psi}_{-20,\downarrow}= -C_- \sin \theta^{\rm
GP},\tilde{\psi}_{20,\downarrow}=A_{\pm}^{\rm GP}$. The power
spectrum in Fig.~3c is after the integration along $r$,
\begin{align}
n_{q,m}\propto \int dr r
|\tilde{\psi}_{q,m}(r)|^2,\sum_{q,m}n_{q,m}=1.
\end{align}

\subsection{Variational calculations}
We consider the SOAMC ground state as either the stripe phase with
$|C_+ C_-|>0$ or the separated phase with $|C_+ C_-|=0$, i.e.,
$C_+=1,C_-=0$ or $C_-=1,C_+=0$. The former corresponds to a density
stripe and the latter to no density stripe. In the variational
calculation using two-quasiangular-momentum ansatz where $A_{\pm}$
is absent in the wave function, $0\leq \beta=|C_+|^2 |C_-|^2 \leq
1/4$ and $\beta=1/4$ corresponds to $|C_+|^2=|C_-|^2=|C_+ C_-|=1/2$.
We compare the energy of $\beta=1/4$ and of $\beta=0$, and take the
lower one as the ground state. This is valid because the lowest
energy is at either $\beta=1/4$ or $\beta=0$, i.e., no energy
maximum within $0\leq \beta=|C_+|^2 |C_-|^2 \leq 1/4$. We find this
condition holds by numerically checking the second order derivative
of $\varepsilon^{\rm var0}$, which is negative for $0\leq \beta \leq
1/4$ and $\bar{n}g_1/E_L < 2$. As for the calculation using
four-quasiangular-momentum ansatz for the stripe phase, we compare
the energy $\varepsilon^{\rm var}$ of the stripe phase, $C_{\pm}=
\mp \sqrt{(1-2A_{\pm}^2)/2}$, and of the separated phase with
$C_+=1,C_-=0,A_{\pm}=0$. It is valid to take $C_{\pm}= \mp
\sqrt{(1-2A_{\pm}^2)/2}$ for the stripe phase since the GP results
confirm that this condition holds, which has the time reversal
symmetry, Eq.~\eqref{eq:Cpm_Apm}.

\subsection{Trap parameters of the simulations}
We indicate the trap parameters: for data in Fig.~3 with $\mu= h
\times 21$~Hz and $R_{\rm TF}=46\micron$,
$N=10^4,\omega_r/2\pi=1.5$~Hz, and $\omega_z/2\pi=600$~Hz. For data
in Fig.~4 with $\mu= h \times 93$~Hz and $R_{\rm
TF}=12.5,25,50\micron$, $N=0.25\times 10^4,10^4, 4\times 10^4,
\omega_r/2\pi=11.744,5.872, 2.936$~Hz respectively;
$\omega_z/2\pi=1000$~Hz.

\subsection{Comparison to SLMC systems}
We list results of spin-linear-momentum coupled (SLMC) BECs from
Ref.~\cite{Li2012}. Two counter-propagating Raman beams along $x$
transfer linear momentum $\Delta k_x=2k_r$ between spin
$\ket{\uparrow}$ and $\ket{\downarrow}$, producing SLMC. The linear
momentum transfer $2k_r$ is analogous to the OAM transfer $\Delta
\ell$ in SOAMC, and thus $4E_r$ is equivalent to $E_L$ in SOAMC. A
spinor wave function ansatz analogous to our
two-quasiangular-momentum ansatz, Eq.~\eqref{eq:spinor_Psi}, is
employed. For a uniform system with no trapping potentials, the
critical coupling is
\begin{align}
\hbar\Omega_c^{\rm SLMC}\approx \sqrt{\frac{2g_2}{g_1}}4E_r
\sqrt{1+\frac{\bar{n}g_1}{4E_r}}.
\end{align}
after expanding to first order in $\bar{n}g_1/4E_r$ for small
interaction $\bar{n}g_1/4E_r$. The stripe contrast is
\begin{align}
\eta^{\rm SLMC}=\frac{\hbar \Omega}{4E_r(1+\bar{n}g_1/4E_r)},
\end{align}
and is
\begin{align}
\eta^{\rm SLMC}\approx \frac{\hbar
\Omega}{4E_r}(1-\frac{\bar{n}g_1}{4E_r})
\end{align}
after expanding to first order in $\bar{n}g_1/4E_r$, which is the
same as $\eta^{\rm var0}$ in Eq.~\eqref{eq:theta_var0_expand} based
on ansatz Eq.~\eqref{eq:spinor_Psi}.

\subsection{Scheme of using synthetic clock states}
We propose to use synthetic clock states in the SOAMC system of
$^87$Rb atoms. Here the discussions are based on
Ref~\cite{Trypogeorgos2018}. These clock states are
$|x\rangle,|y\rangle,|z\rangle$, each of which is a
radio-frequency-dressed state, and thus a superposition of bare spin
states $\ket {m_F=0,\pm1}$. The lowest, middle, and highest-energy
dressed state corresponds to $|z\rangle,|x\rangle,|y\rangle$,
respectively. By choosing proper rf parameters, the $xz$ transition
frequency can be made fourth-order sensitive to rf detuning, and
thus to the bias field. We consider a two-level system of
Raman-coupled $|x\rangle$ and $|z\rangle$.

The mean field energy can be expressed in the basis of $|x\rangle$
and $|z\rangle$,
\begin{align}
E_{\rm int}=\int d^3 r \left( \frac{\mathcal{G}_{xx}}{2}|\psi_x|^4+
\frac{\mathcal{G}_{zz}}{2}|\psi_z|^4
+\mathcal{G}_{xz}|\psi_x|^2|\psi_z|^2\right)
\end{align}
with  effective interactions
$\mathcal{G}_{xx},\mathcal{G}_{zz},\mathcal{G}_{xz}$, and
$\mathcal{G}=(\mathcal{G}_{xx}+\mathcal{G}_{zz})/2,\mathcal{G}_1=(\mathcal{G}+\mathcal{G}_{xz})/2,
\mathcal{G}_2=(\mathcal{G}-\mathcal{G}_{xz})/2$. We consider rf Rabi
coupling $\Omega_{\rm rf}=2.77 \omega_q$ at zero detuning where
$\omega_q$ is the quadratic Zeeman energy. This gives
$\mathcal{G}_{xx}=c_0/\sqrt{2\pi}R_z,\mathcal{G}_{zz}=(c_0+0.97
c_2)/\sqrt{2\pi}R_z$, $\mathcal{G}_{xz}=(c_0+0.825
c_2)/\sqrt{2\pi}R_z$, where $c_0=4\pi\hbar^2(a_0+2a_2)/3m$ and
$c_2=4\pi\hbar^2(a_2-a_0)/3m<0$, where $a_f$ is the s-wave
scattering length in the total spin $f$ channel. (Note that
$g_{00}=c_0/(\sqrt{2\pi}R_z)$ and
$g_{-1,-1}=(c_0+c_2)/(\sqrt{2\pi}R_z)$.) The resulting
$\mathcal{G}_1/\mathcal{G}_2\approx 0.17 c_2/c_0$, which is about
$70~\%$ of the $g_2/g_1\approx 0.25 c_2/c_0$. Therefore, the stripe
contrast using the synthetic clock states is similar to our
simulations using bares spin $|0\rangle,|-1\rangle$. If we choose
the two levels as $\ket{x},\ket{y}$ instead,
$\mathcal{G}_1/\mathcal{G}_2\approx 0.35 c_2/c_0$, even bigger than
$g_2/g_1$.

Consider the detuning window within which the stripe phase exist. At
$\Omega_M=0$, the window is $\gtrsim 1$~Hz for our data in Fig.~1c,
where $\mu=c_0 n_{\rm 3D}=h\times 926$~Hz, $c_2 n_{\rm 3D}=h\times
4.2$~Hz, and $n_{\rm 3D}$ is the peak 3D density. This can be
potentially observed given the measured stability of $\sim
0.1-1$~Hz.

\subsection{Validity of the symmetric inter-spin interaction}
We verify the stripe phases with the realistic $g_{\uparrow
\uparrow}\neq g_{\downarrow \downarrow}$ are approximately the same
as that with symmetric inter-spin interaction, $g_{\uparrow
\uparrow}= g_{\downarrow \downarrow}=g$, while with a detuning
shift. We obtain the phase diagram using realistic $g_{\uparrow
\uparrow}, g_{\downarrow \downarrow}$, and identify the ground state
with the maximum stripe contrast is at $\delta/2\pi=1.4$~Hz, instead
of $\delta=0$ for $g_{\uparrow \uparrow}= g_{\downarrow
\downarrow}=g$. The parameters are $\Delta \ell=40,
r_M=4.25\micron,R_{\rm TF}=10\micron$, $N=1.55 \times 10^4,
\omega_r/2\pi=45.746$~Hz, and $\omega_z/2\pi=1000$~Hz. The critical
coupling is $\Omega_c/2\pi=731.7$~Hz where the peak contrast is
$\eta^{\rm GP}=0.217$. This is very close to that with $g_{\uparrow
\uparrow}= g_{\downarrow \downarrow}=g$, where
$\Omega_c/2\pi=730.0$~Hz and $\eta^{\rm GP}=0.210$.

\section{Acknowledgements}
Y.~-J.L. was supported by MOST and Thematic Program in Academia
Sinica. Y. K. was supported by JST-CREST (Grant No. JPMJCR16F2) and
JSPS KAKENHI (Grants No. JP18K03538 and No. JP19H01824). S.~-K.Y.
was supported by MOST Grant number 107-2112-M001-035-MY3.

\bibliography{SOAMC_phases}
\end{document}